\documentclass[]{llncs}

\usepackage{makeidx}  

\usepackage{latexsym}
\usepackage[pdftex]{graphics}
\usepackage[all,v2,cmtip,2cell]{xy}
\UseAllTwocells
\usepackage{xypic}
\usepackage{amsmath}
\usepackage{amssymb}
\xyoption{all} 
\xyoption{rotate}
\xyoption{v2}
\usepackage{rotating}
\usepackage{mathrsfs}
\usepackage{eufrak}
\usepackage{color}

\usepackage{tikz-cd,mathtools}
\tikzset{mydescription/.style={anchor=center,fill=white}}

\newcommand{\Zp}{{\mathbb Z_p}}

\begin{document}

\author{Peter M. Hines}
 
 \institute{University of York}
 
 \title{A diagrammatic approach to information flow in encrypted communication (extended version)}

\maketitle
\begin{abstract}{\em We give diagrammatic tools to reason about information flow within encrypted communication. In particular, we are interested in deducing where information flow (communication or otherwise) has taken place, and fully accounting for all possible paths. 

The core mathematical concept is using a single categorical diagram to model the underlying mathematics, the epistemic knowledge of the participants, and (implicitly) the potential or actual communication between participants.
A key part of this is a `correctness' or `consistency' criterion that ensures we accurately \& fully account for the distinct routes by which information may come to be known (i.e. communication and / or calculation).  

We demonstrate how this formalism may be applied to answer questions about communication scenarios where we have the partial information
about the participants and their interactions. Similarly, we show how to analyse the consequences of changes to protocols or communications, and to enumerate the distinct orders in which events may have occurred.

We use various forms of Diffie-Hellman key exchange as an illustration of these techniques. However, they are entirely general; we illustrate in an appendix how other protocols from non-commutative cryptography may be analysed in the same manner.}
\end{abstract}

\section{Introduction}\label{intro} 
This paper is about using categorical diagrams to study (or rather, reconstruct) the flow of information in scenarios involving encrypted communication; it is not about the difficulty or otherwise of solving mathematical problems on which security is based. 

\subsection{Key aims}
The main aim of this paper is to introduce tools, based on diagrammatic representations, to ensure that we have fully accounted for {\em information flow} and {\em routes to calculating values} in communication generally, and cryptographic protocols specifically.  Of course, we do not expect to find within the cryptographic literature examples  of protocols where the designers have failed to do this! Rather, we use pre-existing protocols (in particular various forms of Diffie-Hellman key exchange, as a particularly well-understood example) in order to motivated and test our tools.

The utility comes when we use such techniques to reason about incomplete, rather than complete, descriptions of encrypted communication.  We wish to analyse the situation where -- for example -- one participant becomes aware of some additional information (say, a secret key). Can we deduce, in a systematic way, the additional routes to calculating values that this implies? This is considered in Section \ref{leakybob}.

Alternatively, we wish to work backwards -- we know that some private information has become more widely known, but no single individual is in a position to have shared this. Can we fully account for all possible routes by which this information became known? This is considered in Section \ref{incomplete-sect}. 

\subsection{Tools used}
Our starting point is the common category-theoretic technique of expressing algebraic identities via commuting diagrams.  
Drawing such diagrams for the algebra behind cryptographic protocols makes their structure clear (see, for example \cite{DP}), and gives a clear representation of the underlying mathematics; this paper also extends such diagrams to include the participants, and what they come to know in the course of the protocol. 

Mathematically, we do this by moving beyond commuting diagrams, and recovering both the information flow between participants, and distinct routes by which significant values may be computed,  as 2-categorical structure. 

Based on this, we give a `correctness' criterion that ensures that potential or actual information flow within the diagram is modelled correctly -- i.e. nothing has been `left out' and we have not overlooked any route by which a participant may come to know some information.



\section{Bipartite Diffie-Hellman, diagramatically}\label{DHsect}
We use, as illustration, the basic bipartite D-H protocol  \cite{DH,RM}; the underlying algebra, communications, and knowledge of participants are we summarised in Table \ref{DHtable}.

\begin{table}\caption{A concise summary of D-H key exchange}\label{DHtable}
\begin{center}
\begin{tabular}{lcr}
\scalebox{0.8}{
\begin{tabular}{|l|c|r|} 
\hline
{\bf Alice} & {\bf  Public}	& {\bf Bob} \\ 
\hline\hline
												& Public prime $ p$ &	\\
												& Public root $ g\in \mathbb Z_p$ & \\ 
												\hline \hline 
Selects  private  									&	& Selects  private  \\
${ a} \in { \mathbb Z_p}$										&	& ${ b} \in { \mathbb Z_p}$ \\
\hline
\hline
Computes ${ g}^{ a}$		& $\xymatrix{  \ar[rr]^{\mbox{ Announces } g^a} && }$ 	&	\\
\hline
																& $\xymatrix{  && \ar[ll]_{\mbox{ Announces }  {g^b}}}$	& Computes ${ g}^{ b}$ \\
\hline
\hline
Computes: ${ \left( g^b\right)}^{a} $ 			& 	& Computes:  ${ \left( g^a\right)}^{b} $  \\
\hline 
\multicolumn{3}{|c|}{\em By elementary arithmetic, these are equal.}  \\
\multicolumn{3}{|c|} {${ \left( g^b\right)}^{a} \  =\  {g^{ab}} \ = \ { \left( g^a\right)}^{b} $} \\ 
\hline
\end{tabular}
}
&
\ \ \  
&
\scalebox{0.7}{
\begin{tabular}{c}
$\xymatrix{ 
				&	*+[F]\txt{{\bf Alice, Bob, Eve} \\ $g$, $g^a$, $g^b$} 									& 				\\
				& *+[F]\txt{{\bf Alice \& Bob} \\ $g^{ab}$ }	\ar@{-}[u]	 \ar@{-}[ld] \ar@{-}[dr]						&							\\
				*+[F]\txt{{\bf Alice}\\ $a$}			&																			& *+[F]\txt{{\bf Bob} \\ $b$} \\
				& *+[F]\txt{{\bf Nobody} \\ $ab$}	\ar@{-}[ul] \ar@{-}[ur]				&						\\
			}$
\end{tabular}
}
\end{tabular}
\end{center}
\end{table}

The tabular presentation simply distinguishes {\bf public} and {\bf private} information; by contrast, a fine-grained description of the knowledge of the participants (Alice, Bob, and some putative evesdropper\footnote{Although it is standard to assume that Eve is an adversary to Alice and Bob, the tools themselves take a more agnostic approach.  Our aim is to model (or rather, reconstruct) information flow generally; although we are naturally more concerned about information flow to Eve, the mathematical models themselves treat her equally to the other participants.} Eve) is given in lattice form, by `tagging' each algebraic element by a member of the power set lattice $2^{ \{ A,B,E \} }$ of participants. 

\subsection{Expressing algebraic identities diagrammatically}\label{DHdiagrams}
A core category-theoretic practice is giving identities as {\em commuting diagrams}.  
\begin{definition}
A {\bf diagram} over a category $\mathcal C$ is simply a directed graph with nodes labeled by objects. Each edge is labeled by an arrow whose source / target is given by the labels on the initial / final nodes. A diagram {\bf commutes} when the composites along all paths with the same starting / finishing node are equal.  
\end{definition}

\begin{figure}[h]\caption{Bipartite Diffie-Hellman key exchange}\label{DHdiagram}
		\begin{center}
	\scalebox{0.65}{
	$ \xymatrix{
		&&			& \mathbb Z _P \ar[ddrrr]^{(\_ )^b  }	\ar[ddlll]_{(\_ )^a  }																	&		&&		\\
		&&			&																																&		&&		\\
		\mathbb Z_p\ar[ddrrr]_{(\_ )^b  }		&&			&	\{ * \} \ar[uu]|{[g]  } \ar[rrr]|{\left[ g^b\right]  }	 \ar[lll]|{\left[ g^a\right]  } \ar@[red][dd]|{\color{red}\left[ g^{ab} \right] } &		&&	\mathbb Z_p	\ar[ddlll]^{(\_ )^a  }	 \\
		&&			&																																&		&&				\\
		&&			& \mathbb Z_p																													&		&&		
	}
	$ 
}
\end{center}			
\end{figure}
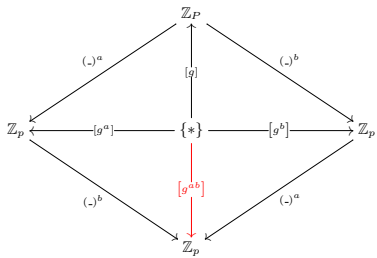

Although the concept is simple, commuting diagrams provide a very efficient and visual way to express algebraic identities.  In Figure \ref{DHdiagram} we express the identies from Table \ref{DHtable} as a commuting diagram over the following category:
\begin{definition}\label{DHp}
	Given prime $p\in {\mathbb N}$, we define the category $\bf DH_p$ to have two objects: a singleton object $\{ * \}$ and the set $\mathbb Z_p=\{ 0,\ldots,p-1\}$. 

For all $x=0,\ldots,p-1$, we have the following arrows:
		\begin{itemize}
			\item The {\bf selection} arrows $[x]:\{* \} \rightarrow \mathbb Z_p$, defined by $[x](*) = x\in \mathbb Z_p$. 
			\item The {\bf modular exponentiation} arrows $(\_ )^x:\mathbb Z_p\rightarrow \mathbb Z_p$, defined in the usual arithmetic manner.
		\end{itemize}
\end{definition}

\begin{remark}[Interpretation]  The arrows of the above category should be though of as {\em operations} that may reliably be {\em performed by participants}.  When we say that ``Bob knows $g^{ab}$'', we mean that Bob is able to select $g^{ab}$ from the whole of $\mathbb Z_p$.  This is the interpretation of the `selection arrows'.
\end{remark}

\subsection{Combining algebraic \& epistemic data}\label{AE}
\begin{figure} \caption{The Algebraic-Epistemic diagram for Diffie-Hellman key exchange}\label{DH_AE-fig}
\begin{center}\scalebox{0.8}{
$\xymatrix{
		&&			& \mathbb Z _P \ar[ddrrr]^{(\_ )^b , \{ B\} }	\ar[ddlll]_{(\_ )^a , \{ A\} }																	&		&&		\\
		&&			&																																&		&&		\\
		\mathbb Z_p\ar[ddrrr]_{(\_ )^b , \{ B\} }		&&			&	\{ * \} \ar[uu]|{[g] , \{ A,B,E\} } \ar[rrr]|{\left[ g^b\right] , \{ A,B,E\} }	 \ar[lll]|{\left[ g^a\right] , \{ A,B,E\} } \ar[dd]|{\left[ g^{ab} \right], \{ A,B \} } &		&&	\mathbb Z_p	\ar[ddlll]^{(\_ )^a , \{ A\} }	 \\
		&&			&																																&		&&				\\
		&&			& \mathbb Z_p																													&		&&		
	}
$
}
\end{center}	
\end{figure}
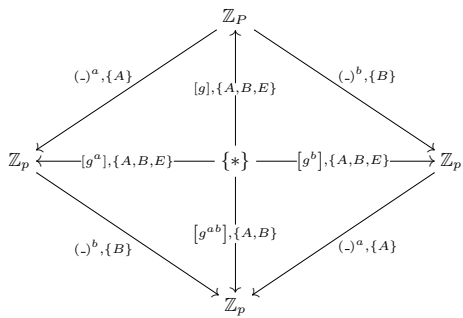

We now combine the algebraic and epistemic aspects of the D-H protocol into a single categorical diagram (Figure \ref{DH_AE-fig}), by `tagging' each arrow by the subset of participants that are able to perform that operation. 

By treating $2^{\{ A,B,E \} }$ as a monoid with composition given by intersection we consider Figure \ref{DH_AE-fig} to be a categorical diagram over the product category ${\bf DH_p}\times 2^{\{ A, B , E\} }$.  Note that this categorical diagram {\em fails to commute}.   We discuss the significance of this in Section \ref{comfail-sect} below, but first provide some much-needed clarification on what, precisely, is being modeled by these diagrams.

\subsection{What is being modeled by  A-E diagrams?}\label{explain-sect}
The interpretation of the epistemic tags in an A-E  is that an algebraic component (e.g. a private key, public message, shared secret, etc.) is labeled by some representation of `who knows this value'.  What has not been included is any explicit representation of {\em how} or {\em when} they came to know this.  There is therefore no in-built notion of `event', `message',  `time-ordering', `causality' or even `communication'.

Rather,  the information in the diagram simply tells us that a given participant became aware of a given value {\em at some point}. It may be seen as a retrospective view of some communication protocol.  Once the process has completed, who has become aware of what?

This is intentional, and by design.  One aim is to demonstrate how, given this restricted information, we may nevertheless reconstruct possible scenarios of how this state of affairs may have arisen (via communications between participants, and participants using the result of these communications to calculate new values).  

For a well-designed cryptographic protocol, we expect this to be unique, or at least unique up to some inessential re-ordering of events (such as well-known variations of the steps in tri-partite Diffie-Hellman key exchange of Section \ref{3-3-DH}).   

Of more interest is the situation where there is some ambiguity, or simply where something has gone wrong! We may wish account for all routes by which some information became public knowledge (Section \ref{incomplete-sect}), or to analyse the consequences of some individual having more a priori knowledge than we had anticipated (Section \ref{leakybob}).

\section{Information flow as failure of commutativity}\label{comfail-sect}
The failure of commutativity in Figure \ref{DH_AE-fig} is obvious. Our claim is that this is a feature rather than a bug:  non-trivial information flow becomes obvious in this graphical form. Precisely, the points at which commutativity fails are those where either 1/ a public announcement has taken place, or 2/ there exists more than one route to calculating the same result. 

\begin{figure} \caption{Announcements as failure of commutativity in  D-H key exchange}\label{DH-pub}
\begin{center}
\scalebox{0.8}{ \xymatrix{
		\framed{ \txt{Alice's \\ announcement}}		&&						& \mathbb Z _P \ar[ddrrr]^{(\_ )^b , \{ B\} }	\ar[ddlll]_{(\_ )^a , \{ A\} }	\ar@{}[ddl]^(.35){}="a"^(.65){}="b" \ar@{=>} "a";"b"			\ar@{}[ddr]^(.35){}="c"^(.65){}="d" \ar@{=>} "c";"d"												&		&&	\framed{\txt{Bob's \\ announcement}}		\\
		&&						&																																&		&&		\\
		\mathbb Z_p								&&						&	\{ * \} \ar[uu]|<<<<{[g] , \{ A,B,E\} } \ar[rrr]|{\left[ g^b\right] , \{ A,B,E\} }	 \ar[lll]|{\left[ g^a\right] , \{ A,B,E\} } ="apub"			 &		&&	\mathbb Z_p		 \\
		&&						&																																&		&&					\\
	}
}
\end{center}
\end{figure}
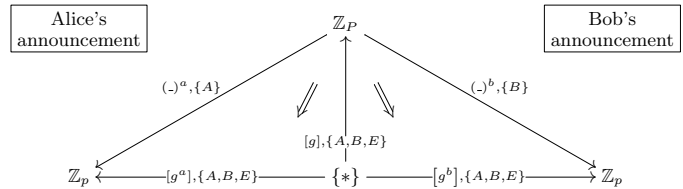

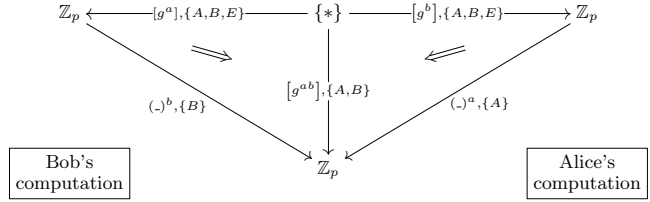
\begin{figure} \caption{Failure of commutativity via distinct paths to the same result}\label{DH-comp}
\begin{center}
\scalebox{0.8}{\xymatrix{
		\mathbb Z_p\ar[ddrrr]_{(\_ )^b , \{ B\} } \ar@{}[drrr]^(.45){}="a"^(.65){}="b" \ar@{=>} "a";"b"		&&			&	\{ * \}  \ar[rrr]|{\left[ g^b\right] , \{ A,B,E\} }	 \ar[lll]|{\left[ g^a\right] , \{ A,B,E\} } \ar[dd]|{\left[ g^{ab} \right], \{ A,B \} } &		&&	\mathbb Z_p	\ar[ddlll]^{(\_ )^a , \{ A\} }	 \ar@{}[dlll]^(.45){}="c"^(.65){}="d" \ar@{=>} "c";"d"	 \\
		&&			&																																&		&&				\\
		\framed{ \txt{Bob's \\ computation}}			&&			& \mathbb Z_p																													&		&&		\framed{ \txt{Alice's \\ computation}}
	}
}
\end{center}			
\end{figure}

Consider the subdiagram of Figure \ref{DH_AE-fig} given in Figure \ref{DH-pub}. This fails to commute because 
$\left( (\_ )^a ,\{ A\} \right) \left( [g], \{ A,B,E \} \right) =   \left(  \left[ g^a\right] , \{ A \} \right)   \neq  \left(  \left[ g^a\right] , \{ A,B,E \} \right)$. 
Similarly, $ \left( (\_ )^b ,\{ B\} \right) \left( [g], \{ A,B,E \} \right)  =   \left(  \left[ g^b\right] , \{ B \} \right)   \neq  \left(  \left[ g^b\right] , \{ A,B,E \} \right) $. 

The underlying cause in both cases is the public announcements: we would see the label $\left(  \left[ g^a\right] , \{ A \} \right) = \left( (\_ )^a ,\{ A\} \right) \left( [g], \{ A,B,E \} \right)$ in the case where Alice had raised the publicly known root to her secret key, but {\em kept the result to herself}. Similarly, we see an edge labeled by $\left(  \left[ g^b\right] , \{ A,B,E \} \right)$, rather than $\left(  \left[ g^b\right] , \{ B \} \right) =  \left( (\_ )^b ,\{ B\} \right) \left( [g], \{ A,B,E \} \right)$ because Bob has {\em publicly shared the result of his computation}. 

Communication between participants clearly causes failure of commutativity; however, there is another significant reason why a diagram may fail to commute. 
Figure \ref{DH-comp} gives another subdiagram of Figure \ref{DH_AE-fig} that also fails to commute, since 
$\left(  (\_ )^b , \{ B \} \right) \left( [g^a ] , \{ A,B,E \} \right)  = \left( [ g^{ab} ] , \{ B \} \right)  \neq   \left( [ g^{ab} ] , \{ A,B \} \right)$. 
In a similar way, $\left(  (\_ )^a , \{ A \} \right) \left( [g^b ] , \{ A,B,E \} \right) = \left( [ g^{ab} ] , \{ A \} \right) \ \neq  \ \left( [ g^{ab} ] , \{ A,B \} \right)$. 

However, no announcements or sharing of information have taken place in this part of the protocol. Rather commutativity fails because Alice and Bob have separately arrived at the same information (i.e. their shared secret $g^{ab}$) via two distinct paths.  

The fact that they both know it (and only they know it) is accounted for by the fact that the label on shared secret is the {\em join} of the labels of the two paths with the same source and target. 
 

\section{Algebraic-Epistemic diagrams, and a correctness condition}
The above considerations apply generally, and motivate the following definitions:
\begin{definition}
We define the {\bf Algebraic-Epistemic} or {\bf A-E diagram} for a communication protocol or scenario to be a categorical diagram giving a complete representation of thealgebraic components, together with tags representing who becomes aware of what information. 
\end{definition} 

\begin{remark} This paper uses Diffie-Hellman key exchange as illustration because it is simple and well-understood.  We emphasise that these techniques are general;  we may draw similar diagrams for other communication protocols or scenarios. Appendix \ref{CAKEsection} gives A-E diagrams for the `Commuting Action Key Exchange' family of protocols from non-commutative cryptography \cite{SZ}, and demonstrates that the same interpretations and correctness criteria hold.  
\end{remark}

\subsection{A correctness criterion for A-E diagrams}
We now introduce a general `correctness' criterion on A-E diagrams.  This is based on  partial order enriched categories.  

\begin{definition}
A category $\mathcal C$ is {\bf poset--enriched} when each homset $\mathcal C(X,Y)$ has a partial ordering $\leq_{XY}$ compatible with composition:
\[ f\leq_{XY} a\in \mathcal C(X,Y) \ \mbox{ and } g \leq_{YZ} b \in \mathcal C(Y,Z) \ \Rightarrow \ gf \leq_{XZ} ba  \in \mathcal C(X,Z) \]
(It is common to omit the object subscripts; these are generally clear from the context).

Any category may be considered to be enriched over the partial order given by equality on homsets.  The product of two poset-enriched categories is also assumed to be enriched over the product partial order: $(a,b) \leq (c,d)$ iff $a\leq c$ and $b \leq d$.   Thus we may assume the category ${\bf DH_p}\times 2^{\{ A, B , E \} }$ used in Section \ref{AE} to be poset-enriched.
\end{definition}  
Based on this, we give a general definition on diagrams over poset--enriched categories that we will claim as a `correctness criterion' for Algebraic-Epistemic diagrams.

\begin{definition}
We say that a diagram $\mathfrak D$ over a poset-enriched category $\mathcal C$ satisfies the {\bf information flow ordering (IFO) condition}, or is an {\bf information flow ordered diagram} when: 
\begin{enumerate}
\item The underlying diagraph of $\mathfrak D$ is acyclic.
\item 
For any edge $e$ and path $p=p_k\ldots p_1$ with the same source and target node, the label on $p$ is $\leq$  the label on $e$. 
\end{enumerate}
\end{definition}

\begin{remark}[The general setting]  Poset-enriched categories are a very special case of 2-categories, where as well as objects and arrows between objects, we have `higher-level' notion of 2-morphisms between arrows. We refer to \cite{JP} for a good exposition of the general theory, but this is only relevant for our purposes in that we have a neat ready-made diagrammatic calculus.

It is standard to draw 2-morphisms in categorical diagrams as ``two-cells''; for our purposes these are simply labeled by the partial order relation, so condition 2. is drawn as follows:  
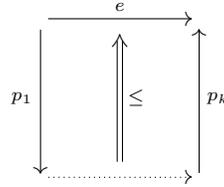
\[ 
\xymatrix{
\ar[rr]^{e}_{\ }="2target" \ar[dd]_{p_1} 	&					&		\\
							&					&		\\
\ar@{.>}[rr]^{\ }="2source"			&			 		&	\ar[uu]_{p_k}	\ar@{=>} "2source" ; "2target"  _{\leq}  \\
}
\]
An immediate consequence of this condition is that any pair of edges with the same source and target nodes have the same label. We therefore include the assumption there is at most one edge with a given source / target.
\end{remark}

\begin{remark}[The IFO condition as a correctness criterion]
The IFO condition is proposed as a correctness criterion for Algebraic-Epistemic diagrams generally. This `correctness'  is simply about about accurately accounting for 1/ information flow between participants, and 2/ what this enables them to calculate.   Our claim is that if we find that the IFO condition is not satisfied, we have failed to account for either 1/ or 2/.  Further, we may often recover this additional information in a systematic and easily automated manner.
\end{remark}

\subsection{Justifying the IFO condition}\label{EPexplained}  
The prescription for drawing A-E diagrams is entirely generic. 
Diagrams are drawn over a category of the form $\mathcal C \times \mathcal L$, where $\mathcal C$ is the algebraic setting for the protocol, and $\mathcal L$ is a meet-semilattice (generally the powerset-lattice $2^P$ of the participants in the protocol).  We assume $\mathcal C$ to be poset-enriched over the equality relation, so the product category $\mathcal C\times 2^P$ is then enriched via the product partial ordering.

The Algebraic-Epistemic diagram $\mathfrak D$ for a protocol is a diagram over this category. The projection onto the first component $\pi_1 \left( \mathfrak D \right)$ is an acyclic commuting diagram over $\mathcal C$ that expresses the relationships between operations performed by participants in the protocol.  By construction, this commutes, and  therefore trivially satisfies the IFO condition. The additional lattice labels in $\mathfrak D$ itself are `tags' giving the subset of participants that are able to perform the operation on that edge.  

Based on this generic description, the interpretation of the IFO condition is straightforward. 
Consider (a fragment of) the A-E diagram for some protocol consisting of one edge and one path between nodes $H$ and $K$, as follows:
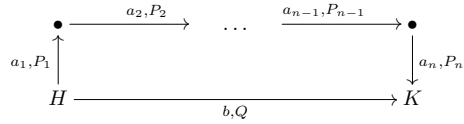
\[ \scalebox{0.8}{\xymatrix{ 
\bullet \ar[rr]^{a_2,P_2} 				&&\ \ \ \  \ldots \ \ \ \  \ar[rr]^{a_{n-1}, P_{n-1} } 	& & \bullet \ar[d]^{a_n,P_n} \\
H \ar[u]^{a_1,P_1} \ar[rrrr]_{b,Q} 	&&							  	& & K				
}
}
\]
The IFO condition  in this simple case states that $\bigwedge_{j=1}^n P_j \ \leq \ Q$.  Given our interpretation, the IFO condition is an axiomatisation of  the triviality that any individual who is able to perform each of the operations $a_1,\ldots ,a_n$ is also able to perform their composite $a_na_{n-1}\ldots a_1$.  

Conversely, consider some diagram consisting of a single edge from node $H$ to node $K$, and multiple paths $\{ \Pi_1,\ldots \Pi_n\} $with the same source and target, where the meet of the labels along $\Pi_k$ is denoted $R_k$,  as follows:
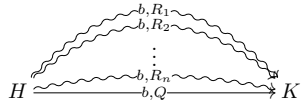
\[ \scalebox{0.8}{\xymatrix{
									&&&	&	\\
H \ar[rrrr]| {b,Q} \ar@{~>}@/^39pt/[rrrr]|{b,R_1}   \ar@{~>}@/^32pt/[rrrr]|{b,R_2}="b"  \ar@{~>}@/^8pt/[rrrr]|{b,R_n}="a"  &&&& K  \\
   \ar @{} "a";"b" |{\vdots}
   }
}
\]
The interpretation of the IFO condition is again straightforward. Every member of $R_1,R_2,\ldots ,R_n$ is able to perform $b$; thus $R_j \leq Q$ for all $j=1 .. n$. Using the additional lattice operations of $2^P$ 
we may also write this as $ \bigvee_{j=1}^n R_j \ \leq \ Q$. However, the possibility that additional communication / announcements have also taken place prevents us from writing $ \bigvee_{j=1}^n R_j \ =  \ Q$; indeed, failure of this condition is a clear signal that additional communication has taken place.

\begin{remark}[The IFO condition and deadlock-freeness]\label{DFclaim}
A further consequence of the IFO condition is {\em deadlock-freeness}; for example, it rules out the situation where Alice is waiting for a communication from Bob before she may continue, whilst simultaneously, Bob is waiting for a communication from Alice before he may take his next step. 

This is not ruled out by the acyclicity of the underlying graph; communication appears as arrows between the edges of this graph (the partial order relations, drawn as  2-cells) -- it is these that we need to ensure do not form closed loops. 

Fortunately, this follows for free from the IFO condition; deadlock would appear as a `closed loop' of inequalities of distinct labels on edges, such as $a<b$, $b<c$, and $c<a$.  However, the anti-symmetry axiom  $x\leq y \ \& \ y \leq x\ \Rightarrow \ x=y$ for partial orders then implies that $a=b=c$.  This contradicts the assumption that $a$, $b$ and $c$ are {\em distinct} labels!
\end{remark}


 \section{Tripartite Diffie-Hellman key exchange}\label{3-3-DH}
We now use diagrammatic methods to compare and contrast two approaches to tripartite secret sharing based on Diffie-Hellman key exchange.
Multi-partite generalisations of Diffie-Hellman key exchange are well-established (see, for example, \cite{MOV}). We consider the case where three participants construct a {\em single shared secret}, and  where each pair of the three participants has a {\em distinct shared secret}.  
We refer to these as $\binom{3}{3}$ Diffie-Hellman and $\binom{3}{2}$ Diffie-Hellman respectively. 

They are of course special cases of the situation where there are $n$ participants, and each subset of $k$ participants constructs a distinct shared secret -- what we refer to as the general $\binom{n}{k}$ Diffie-Hellman protocol. This, including its diagrammatics, is considered in Appendix \ref{3dshapes}.

\begin{definition}[$\binom{3}{3}$ Diffie-Hellman key exchange]\label{3dh}
Let us assume participants $\{ Alice, Bob ,Carol, Eve \}$ where Eve is the evesdropper,  and Alice, Bob, and Carol will construct a mutual shared secret. Alice, Bob and Carol choose private keys $a,b,c\in \mathbb Z_p$ respectively, and their shared secret $g^{abc}=g^{bca}=g^{cab}$ is constructed as follows:

\begin{enumerate}
\item Alice computes $g^a$ and communicates the result to Bob.
\item Bob computes $g^b$ and communicates the result to Carol.
\item Carol computes $g^c$ and communicates the result to Alice.$ $ \\
\item Alice computes $\left( g^c\right)^a = g^{ca}$ and communicates the result to Bob.
\item Bob computes $\left( g^a\right) ^b = g^{ab}$ and communicates the result to Carol.
\item Carol computes $\left( g^b\right) ^c = g^{bc}$ and communicates the result to Alice.$ $ \\
\item Alice computes $\left( g^{bc}\right)^a = g^{abc}$.
\item Bob computes $\left(g^{ca}\right)^b =g^{abc}$
\item Carol computes $\left( g^{ab}\right) ^c = g^{abc}$.
\end{enumerate}
It is of course assumed that Eve is party to all communication. We have made a slight break with convention, simply in order to test the formalism, and assumed that for whatever reason, Carol is not party to the communications between Alice and Bob, etc.   
\end{definition}

The Algebraic-Epistemic diagram for this is given in Figure \ref{trinity}, and -- should it be needed -- a step-by-step description of how this diagram is derived is given in Appendix \ref{stepbystep}. It may be verified that this diagram satisfies the IFO condition, and it is also unambiguous who has communicated what information to whom.
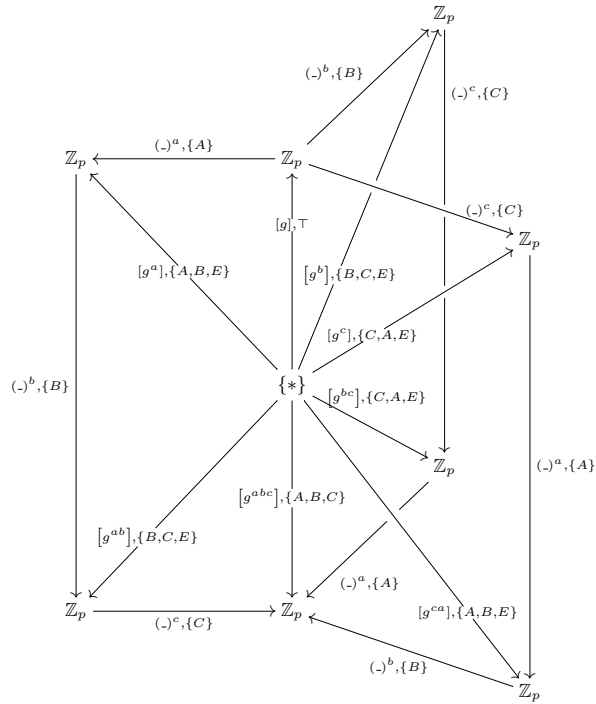
\begin{figure}\caption{Algebraic-Epistemic diagram for $\binom{3}{3}$ Diffie-Hellman}\label{trinity}
\begin{center}
	\scalebox{0.8}
	{
$
\xymatrix{
												&&		&																																																																																																																																					&& \mathbb Z_p\ar[dddddd]^<<<<<<<<<<<{(\_)^c,\{ C \} } |!{[ddll];[dddr]}\hole  |!{[dddddll];[dddr]}\hole			&			\\
												&&		&																																																																																																																																					&& 									 																&			\\
\mathbb Z_p\ar[dddddd]_{(\_)^b,\{ B\} }	&&		&  \mathbb Z_p \ar[uurr]^{(\_ )^b,\{ B\} } \ar[drrr]^>>>>>>{(\_ )^c,\{ C \} } \ar[lll]_{(\_ )^a,\{ A \} }																																																																																																									&&																									&			\\
												&&		&																																																																																																																																					&&																									&\mathbb Z_p 	\ar[dddddd]^{(\_ )^a,\{ A \} }				\\																																				
												&&		&																																																																																																																																					&&																									&		\\
												&&		&	\{ * \} \ar[ddd]|{\left[g^{abc}\right] ,\{ A,B,C\} } \ar[uuu]|>>>>>>>>>{\left[g\right] ,\top}\ar[uuulll]|{\left[g^{a}\right],\{A,B,E\}} \ar[dddlll]|>>>>>>>>>>>>>>{\left[g^{ab}\right],\{ B,C,E \} }	\ar[rruuuuu]|<<<<<<<<<<<<<<<<<{\ \ \ \ \left[g^b\right],\{B,C,E\} }|!{[uuu];[uurrr]}\hole 	\ar[drr]^<<<<<<<{\left[ g^{bc}\right],\{ C,A,E \} }	\ar[uurrr]|<<<<<<<<<<<<{\left[ g^{c}\right], \{ C,A,E \} }	\ar[ddddrrr]|>>>>>>>>>>>>>>{\left[g^{ca}\right], \{ A , B , E \} }	&&																									&			\\
												&&		&																																																																																																																																					&&	\mathbb Z_p \ar[ddll]^>>>>>>>{\left(\_ \right)^a,\{ A \} }|!{[ull];[dddr]}\hole								&								\\						
												&&		&																																																																																																																																					&&																									&			\\																																				
\mathbb Z_p\ar[rrr]_{(\_)^c,\{ C \} }		&&		& \mathbb Z_p																																																																																																																																	&&																									&			\\
												&&		&																																																																																																																																					&&																									&	\mathbb Z_p\ar[ulll]^{(\_ )^b,\{ B \} }	\\
  }
$
}
\end{center}
\end{figure}

An obvious alternative to three participants calculating a single shared secret is the scenario where each pair of participants has a distinct shared secret via the standard Diffie-Hellman protocol.  

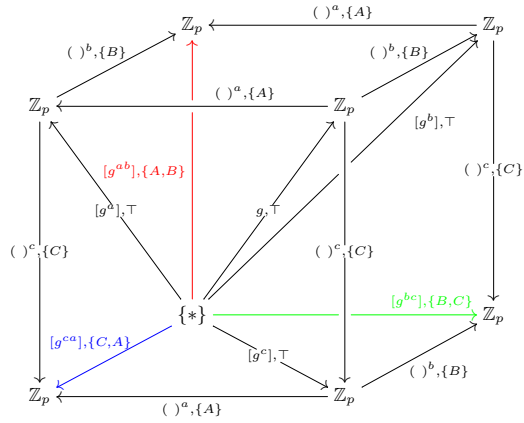
\begin{figure}
\caption{$\binom{3}{2}$ Diffie-Hellman}\label{3-2AE}
\[
\scalebox{0.8}{
\xymatrix{
																					&&	\mathbb Z_p																																																																																																											&&																																		&&	\mathbb Z_p	\ar[dddd]|{( \ )^c, \{ C \} }	\ar[llll]_{(\ )^a, \{ A \} } \\
\mathbb Z_p	\ar[dddd]|{(\ )^c,\{ C \} }	\ar[urr]^{(\ )^b, \{ B \} }			&&																																																																																																															&&	\mathbb Z_p	\ar[urr]^{( \ )^b, \{ B \} } \ar[dddd]|{( \ )^c, \{ C \} } \ar[llll]_<<<<<<<<<<<<<<{(\  )^a, \{ A \} } 		&&			\\
																					&&																																																																																																															&&																																		&&				\\ 
																					&&																																																																																																															&&																																		&&				\\ 
																					&&	\{ * \}	\ar@[red][uuuu]|!{[uuurr];[uuull]}\hole^{\color{red}[g^{ab}], \{ A , B \} }  \ar@[blue][dll]_{\color{blue}[g^{ca}], \{ C ,A \} }	\ar@[green][rrrr]|!{[rru];[rrd]}\hole ^>>>>>>>>{\color{green}[g^{bc}], \{ B,C \} }	\ar[uuurr]|{\color{black}g,\top}	\ar[uuull]|{[g^a],\top }  \ar[drr]|{[g^c],\top } \ar[uuuurrrr]|!{[uuurr];[drr]}\hole _>>>>>>>>>>>>>>>>{[g^b],\top}										&&																																		&&	\mathbb Z_p		\\
\mathbb Z_p																		&&																																																																																																															&&	\mathbb Z_p	\ar[llll]^{( \ )^a, \{ A \} } 	 \ar[urr]	_{( \ )^b, \{ B \} }															&&			\\
}
}
\]
\end{figure}

\begin{definition}\label{3-2DH}
	(The {\bf $\binom{3}{2}$ Diffie Hellman protocol})
We again assume participants $\{ Alice, Bob ,Carol, Eve \}$ where Eve is the evesdropper. Alice, Bob and Carol choose private keys $a,b,c\in \mathbb Z_p$, and each pair, {\color{red}Alice-Bob}, {\color{green}Bob-Carol}, and {\color{blue}Carol-Alice} uses the bipartite D-H protocol to construct a shared secret.  
\begin{itemize}
\item $Alice$, $Bob$, and $Carol$ compute $g^a$ and $g^b$  and $g^c$
respectively. They publicly announce their results. 
\item Alice computes $\color{red}g^{ba}$ (shared with Bob) and $\color{blue}g^{ca}$ (shared with Carol).
\item Bob computes $\color{green}g^{cb}$ (shared with Carol)  and $\color{red}g^{ab}$ (shared with Alice).
\item Carol computes $\color{blue}g^{ac}$ (shared with Alice) and $\color{green}g^{bc}$ (shared with Bob).
 \end{itemize}

\end{definition}
We jump straight to the A-E diagram for the above protocol, given in Figure \ref{3-2AE}. This uses the same colour-coding as above.

\begin{remark}[Ordering of steps in tripartite Diffie-Hellman]\label{ordering3DH}
A notable difference between the step-by-step descriptions of Definitions \ref{3dh} and \ref{3-2DH}, and the A-E diagram of Figures \ref{trinity} and \ref{3-2AE}, is that in the tabular description the order of steps is fixed.  In the categorical diagrams, it becomes clear how this particular ordering of steps is not essential; rather, the only real restrictions are that a participant can only communicate a value {\em after} she has calculated it, and a value can only be calculated once the pre-requisites for this calculation have been received. 

Based on the diagram we may consider alternative orderings of the steps given in Definition \ref{3dh}; it may be verified that these correspond to alternative, but operationally equivalent, presentations multipartite Diffie-Hellman protocols.

\end{remark}

\section{A-E Diagrams as graphical tools for protocols}\label{tools}
Although a diagrammatic approach may give a path to intuitive descriptions of protocols via pictures, we also wish to show how these pictures provide concrete tools for deducing \& reasoning about information flow.  

A diagrammatic calculus allows us easily  to answer certain questions such as, `how much information does a given participant have?', `what are the routes by which an evesdropper may become aware of a given secret?', and `what are the consequences of this particular value becoming known?'.  We first illustrate this using various forms of Diffe-Hellman key exchange, then give general techniques for finding implicit or hidden information via diagrams. 

\subsection{Manipulating A-E diagrams}
We make some straightforward definitions that will have useful interpretations when applied to A-E diagrams.  A key concept is ordering categorical diagrams.

\begin{definition}\label{EPops}
Let $(\mathcal C,\leq)$ be a poset-enriched category, and let  $\mathfrak H, \mathfrak K$ be diagrams (not necessarily commutative) over $\mathcal C$.  We say $\mathfrak H \leq  \mathfrak K$ iff the underlying directed graph of $\mathfrak H$ is a subgraph\footnote{We assume an implicit, fixed, embedding in order not to have to consider the graph embedding or graph isomorphism problem.  In practice, this embedding is immediate from the interpretation}  of the underlying digraph of $\mathfrak K$, and  for all edges of $\mathfrak H$, the label in $\mathfrak H$ is less than or equal to the label of the same edge in $\mathfrak K$.  It is immediate that this a partial order on diagrams over $\mathcal C$. 
\end{definition}

The above is of course applicable to IFO diagrams.  Of particular interest is the poset of IFO diagrams that are above an arbitrary diagram, and whether this poset has a bottom element. In general there may not be a {\em unique} minimal IFO diagram above an arbitrary diagram.  

\subsection{Participants' views of protocols}\label{viewpoints}
A natural example of the ordering of diagrams is given by  taking the A-E diagram for a given protocol, and erasing all edges whose `tag' does not include some participant, or set of participants.  

\begin{figure}\caption{ $\binom{3}{3}$ Diffie-Hellman as seen by various sets of participants}\label{trinity-up}
\scalebox{0.6}
{
\begin{tabular}{ccc}
															&					&															\\
															& \hspace{10em} 	&															\\
{\bf Alice: The $\{ A  \}$ restriction} 	&				 	& {\bf Alice \& Bob: The ${\{ A, B  \} }$ restriction} \\
															&					&																		\\
\xymatrix{
									&&		&																																																																							&& \mathbb Z_p																			&			\\
									&&		&																																																																						&& 						 																&			\\
\mathbb Z_p							&&		&  \mathbb Z_p  \ar[lll]_{(\_ )^a,\{ A \} }																																																														&&																						&			\\
									&&		&																																																																						&&																						&\mathbb Z_p 	\ar[dddddd]^{(\_ )^a,\{ A \} }				\\																																				
									&&		&																																																																						&&																						&		\\
									&&		&	\{ * \} \ar[ddd]|{\left[g^{abc}\right] ,\{ A,B,C\} } \ar[uuu]|>>>>>>>>>{\left[g\right] ,\top}\ar[uuulll]|{\left[g^{a}\right],\{A,B,E\}} 	 	\ar[drr]^<<<<<<<{\left[ g^{bc}\right],\{ C,A,E \} }	\ar[uurrr]|<<<<<<<<<<<<{\left[ g^{c}\right], \{ C,A,E \} }	 \ar[ddddrrr]|>>>>>>>>>>>>>>{\left[g^{ca}\right], \{ A , B , E \} }	&&																						&			\\
									&&		&																																																																						&&	\mathbb Z_p \ar[ddll]^>>>>>>>{\left(\_ \right)^a,\{ A \} }|!{[ull];[dddr]}\hole					&								\\						
									&&		&																																																																						&&																						&			\\																																				
\mathbb Z_p							&&		& \mathbb Z_p																																																																			&&																						&			\\
									&&		&																																																																						&&																						&	\mathbb Z_p	\\
 }
&		& 
\xymatrix{
									&&		&																																														&& \mathbb Z_p							&			\\
									&&		&																																														&& 						 				&			\\
\mathbb Z_p							&&		&  \mathbb Z_p  																																											&&										&			\\
									&&		&																																														&&										&\mathbb Z_p 					\\																																				
									&&		&																																														&&										&		\\
									&&		&	\{ * \} \ar[ddd]|{\left[g^{abc}\right] ,\{ A,B,C\} } \ar[uuu]|>>>>>>>>>{\left[g\right] ,\top}\ar[uuulll]|{\left[g^{a}\right],\{A,B,E\}} \ar[ddddrrr]|>>>>>>>>>>>>>>{\left[g^{ca}\right], \{ A , B , E \} }	 		&&										&			\\
									&&		&																																														&&	\mathbb Z_p 							&								\\						
									&&		&																																														&&										&			\\																																				
\mathbb Z_p							&&		& \mathbb Z_p																																											&&										&			\\
									&&		&																																														&&										&	\mathbb Z_p	\\
 }
\\
																				& \vspace{3em} 	&															\\
{\bf Alice, Bob \& Carol: The ${\{ A,B,C  \} }$ restriction} 	&				 & {\bf Eve: The ${\{ E  \} }$ restriction} \\
																				&				&																		\\

\xymatrix{
									&&		&																															&& \mathbb Z_p							&			\\
									&&		&																															&& 						 				&			\\
\mathbb Z_p							&&		&  \mathbb Z_p  																												&&										&			\\
									&&		&																															&&										&\mathbb Z_p 					\\																																				
									&&		&																															&&										&		\\
									&&		&	\{ * \} \ar[ddd]|{\left[g^{abc}\right] ,\{ A,B,C\} } \ar[uuu]|>>>>>>>>>{\left[g\right] ,\top}								 	 		&&										&			\\
									&&		&																															&&	\mathbb Z_p 							&								\\						
									&&		&																															&&										&			\\																																				
\mathbb Z_p							&&		& \mathbb Z_p																												&&										&			\\
									&&		&																															&&										&	\mathbb Z_p	\\
 }
& 		&		
\xymatrix{
									&&		&																																																																																														&& \mathbb Z_p																			&			\\
									&&		&																																																																																														&& 						 																&			\\
\mathbb Z_p							&&		&  \mathbb Z_p																																																																																											&&																						&			\\
									&&		&																																																																																														&&																						&\mathbb Z_p 				\\																																				
									&&		&																																																																																														&&																						&		\\
									&&		&	\{ * \}  \ar[uuu]|>>>>>>>>>{\left[g\right] ,\top}\ar[uuulll]|{\left[g^{a}\right],\{A,B,E\}} \ar[dddlll]|>>>>>>>>>>>>>>{\left[g^{ab}\right],\{ B,C,E \} }	\ar[rruuuuu]|<<<<<<<<<<<<<<<<<{\ \ \ \ \left[g^b\right],\{B,C,E\} } 	\ar[drr]^<<<<<<<{\left[ g^{bc}\right],\{ C,A,E \} }	\ar[uurrr]|<<<<<<<<<<<<{\left[ g^{c}\right], \{ C,A,E \} }	\ar[ddddrrr]|>>>>>>>>>>>>>>{\left[g^{ca}\right], \{ A , B , E \} }	&&																						&			\\
									&&		&																																																																																														&&	\mathbb Z_p 																			&								\\						
									&&		&																																																																																														&&																						&			\\																																				
\mathbb Z_p							&&		& \mathbb Z_p																																																																																											&&																						&			\\
									&&		&																																																																																														&&																						&		\\
 }
\end{tabular}
 }
\end{figure}
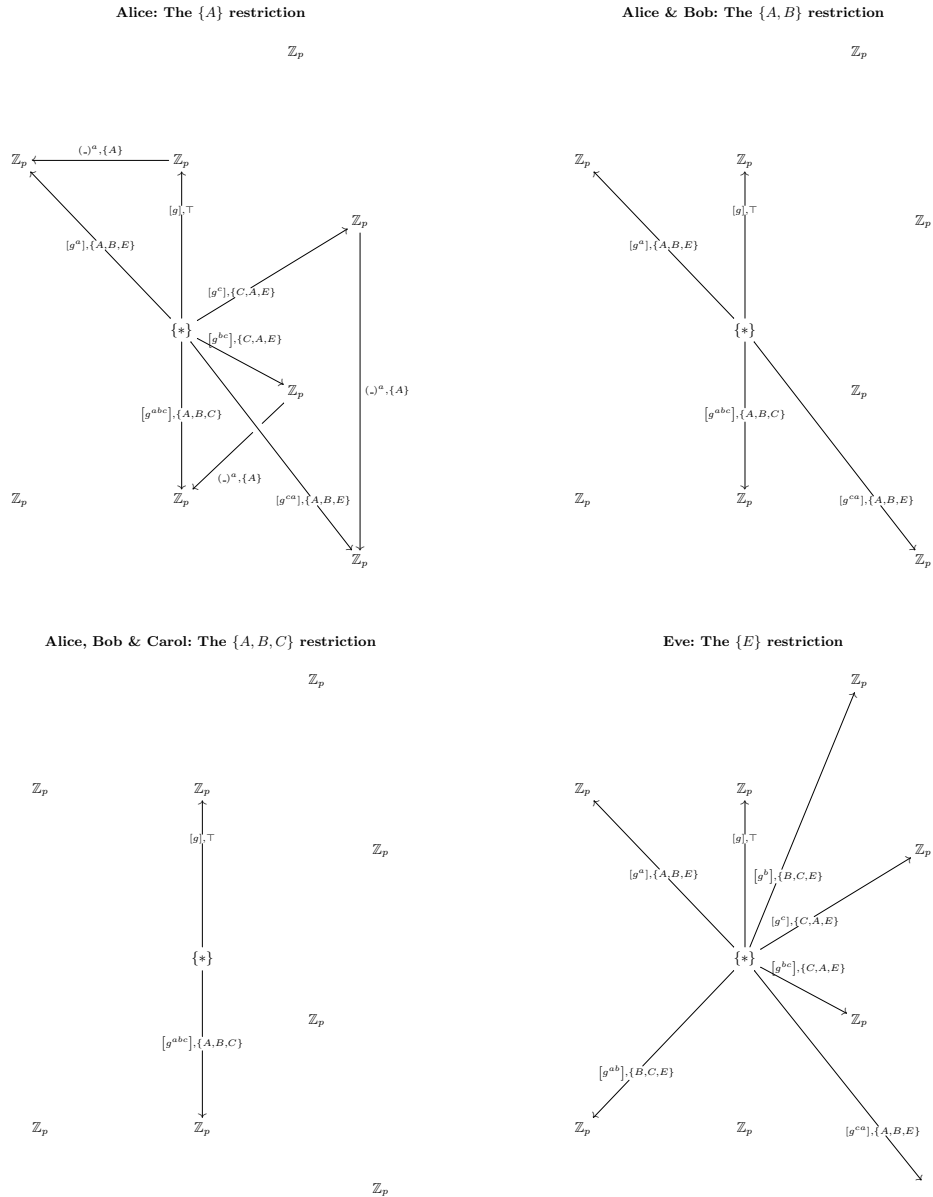

In Figure \ref{trinity-up}, we consider the A-E diagram for the $\binom{3}{3}$ Diffie-Hellman protocol, given in Figure \ref{trinity}, and do this for for the subsets $\{ A\}$, $\{ A,B \}$, $\{ A,B,C \}$ and $\{ E \}$. This gives a convenient graphical illustration of the information available to Alice, Alice and Bob, Alice and Bob and Carol, and the evesdropper respectively. 

It is immediate that these subdiagrams also satisfy the IFO condition, and similarly that taking any A-E diagram satisfying the IFO condition, and erasing all edges according to a simlar criterion, will result in a diagram that again satisfies the IFO condtion. In particular, it is simple to take the diagram of Figure \ref{3-2AE} and erase all edges not accessible to some (non-evesdropper) participant, to recover the A-E diagram for bipartite D-H key exchange given in Figure \ref{DHdiagram}. 

\subsection{Updating A-E diagrams based on additional information}\label{leakybob}
We now consider the more interesting case of when a diagram is modified to reflect some additional information.    The resulting diagram may fail to  satisfy the IFO condition. 

Under these circumstances, the partial ordering of diagrams becomes a useful practical tool: given a diagram $\mathfrak D$ that does not satisfy the IFO condition, we consider the poset of diagrams above it that do satisfy this condition. Under very light assumptions,  this will have a bottom element --- we may analyse this to establish the consequences of this additional information. 

This is best illustrated by a somewhat trivial example; we take both the $\binom{3}{3}$ and the $\binom{3}{2}$ Diffie-Hellman protocols and update them both with some additional information: {\em \bf Eve has become aware of the private key of one of the participants}.  

\begin{figure}\caption{Eve knows Alice's private key!}\label{Eknowsa}
\scalebox{0.6}{
\begin{tabular}{ccc}
															&	&														\\
{\bf Replacing $(\ )^a,\{ A\}$ by $(\ )^a,\{ A,E\}$ }		&	& {\bf The unique smallest IFO diagram above this. }	\\
															&	&														\\
\xymatrix{
												&&		&																																																																																																																																					&& \mathbb Z_p\ar[dddddd]^<<<<<<<<<<<{(\_)^c,\{ C \} } |!{[ddll];[dddr]}\hole  |!{[dddddll];[dddr]}\hole			&			\\
												&&		&																																																																																																																																					&& 									 																&			\\
\mathbb Z_p\ar[dddddd]_{(\_)^b,\{ B\} }	&&		&  \mathbb Z_p \ar[uurr]^{(\_ )^b,\{ B\} } \ar[drrr]^>>>>>>{(\_ )^c,\{ C \} } \ar[lll]_{(\_ )^a,\{ A,E \} }																																																																																																									&&																									&			\\
												&&		&																																																																																																																																					&&																									&\mathbb Z_p 	\ar[dddddd]^{(\_ )^a,\{ A,E \} }				\\																																				
												&&		&																																																																																																																																					&&																									&		\\
												&&		&	\{ * \} \ar[ddd]|{\left[g^{abc}\right] ,\{ A,B,C\} } \ar[uuu]|>>>>>>>>>{\left[g\right] ,\top}\ar[uuulll]|{\left[g^{a}\right],\{A,B,E\}} \ar[dddlll]|>>>>>>>>>>>>>>{\left[g^{ab}\right],\{ B,C,E \} }	\ar[rruuuuu]|<<<<<<<<<<<<<<<<<{\ \ \ \ \left[g^b\right],\{B,C,E\} }|!{[uuu];[uurrr]}\hole 	\ar[drr]^<<<<<<<{\left[ g^{bc}\right],\{ C,A,E \} }	\ar[uurrr]|<<<<<<<<<<<<{\left[ g^{c}\right], \{ C,A,E \} }	\ar[ddddrrr]|>>>>>>>>>>>>>>{\left[g^{ca}\right], \{ A , B , E \} }	&&																									&			\\
												&&		&																																																																																																																																					&&	\mathbb Z_p \ar[ddll]^>>>>>>>{\left(\_ \right)^a,\{ A, E \} }|!{[ull];[dddr]}\hole								&								\\						
												&&		&																																																																																																																																					&&																									&			\\																																				
\mathbb Z_p\ar[rrr]_{(\_)^c,\{ C \} }		&&		& \mathbb Z_p																																																																																																																																	&&																									&			\\
												&&		&																																																																																																																																					&&																									&	\mathbb Z_p\ar[ulll]^{(\_ )^b,\{ B \} }	\\
  }
&		&		
\xymatrix{
												&&		&																																																																																																																																					&& \mathbb Z_p\ar[dddddd]^<<<<<<<<<<<{(\_)^c,\{ C \} } |!{[ddll];[dddr]}\hole  |!{[dddddll];[dddr]}\hole			&			\\
												&&		&																																																																																																																																					&& 									 																&			\\
\mathbb Z_p\ar[dddddd]_{(\_)^b,\{ B\} }	&&		&  \mathbb Z_p \ar[uurr]^{(\_ )^b,\{ B\} } \ar[drrr]^>>>>>>{(\_ )^c,\{ C \} } \ar[lll]_{(\_ )^a,\{ A,E \} }																																																																																																									&&																									&			\\
												&&		&																																																																																																																																					&&																									&\mathbb Z_p 	\ar[dddddd]^{(\_ )^a,\{ A,E \} }				\\																																				
												&&		&																																																																																																																																					&&																									&		\\
												&&		&	\{ * \} \ar[ddd]|{\bf \left[g^{abc}\right] ,\{ A,B,C,E\} } \ar[uuu]|>>>>>>>>>{\left[g\right] ,\top}\ar[uuulll]|{\left[g^{a}\right],\{A,B,E\}} \ar[dddlll]|>>>>>>>>>>>>>>{\left[g^{ab}\right],\{ B,C,E \} }	\ar[rruuuuu]|<<<<<<<<<<<<<<<<<{\ \ \ \ \left[g^b\right],\{B,C,E\} }|!{[uuu];[uurrr]}\hole 	\ar[drr]^<<<<<<<{\left[ g^{bc}\right],\{ C,A,E \} }	\ar[uurrr]|<<<<<<<<<<<<{\left[ g^{c}\right], \{ C,A,E \} }	\ar[ddddrrr]|>>>>>>>>>>>>>>{\left[g^{ca}\right], \{ A , B , E \} }	&&																									&			\\
												&&		&																																																																																																																																					&&	\mathbb Z_p \ar[ddll]^>>>>>>>{\left(\_ \right)^a,\{ A, E \} }|!{[ull];[dddr]}\hole								&								\\						
												&&		&																																																																																																																																					&&																									&			\\																																				
\mathbb Z_p\ar[rrr]_{(\_)^c,\{ C \} }		&&		& \mathbb Z_p																																																																																																																																	&&																									&			\\
												&&		&																																																																																																																																					&&																									&	\mathbb Z_p\ar[ulll]^{(\_ )^b,\{ B \} }	\\
  }

\end{tabular}
}
\end{figure}
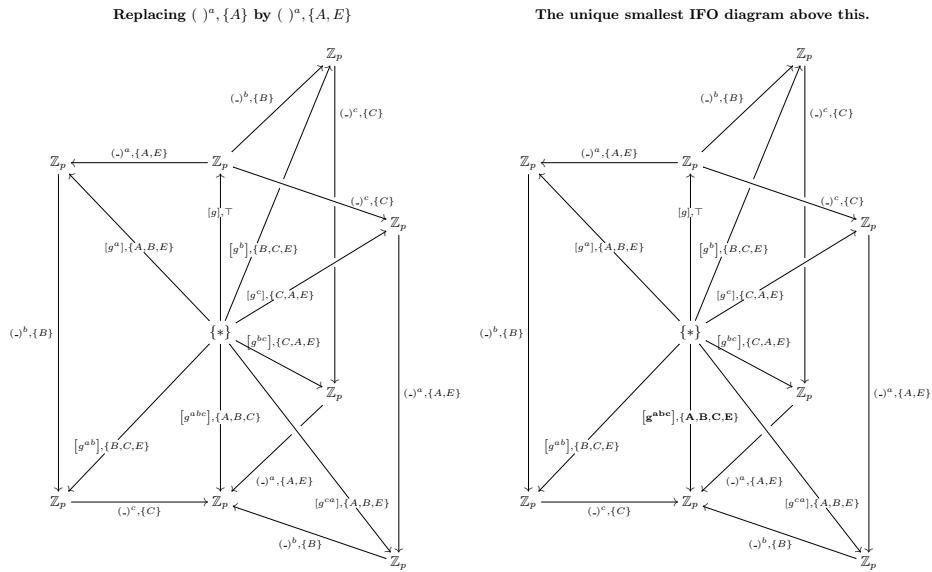

To analyse the $\binom{3}{3}$ protocol,  we modify the diagram of Figure \ref{trinity} to replace every ocurrence of $( \ )^a , \{ A \}$ by $( \ )^a , \{ A, E \}$.  This will result in the diagram on the lhs of Figure \ref{Eknowsa}.
 
This diagram does {\em not} satisfy the IFO condition; it is missing either some communication or some route to participants calculating a given value.  Fortunately, the poset of IFO diagrams above this has a smallest element:  given on the rhs of Figure \ref{Eknowsa}. 

This particular case is straightforward; the lhs diagram has failed to satisfy the IFO condition because of the following single subdiagram:
\[ 
\scalebox{0.8}{
\xymatrix{
	\{ * \} \ar[ddd]|{\left[g^{abc}\right] ,\{ A,B,C\} } 	\ar[drr]^<<<<<<<{\left[ g^{bc}\right],\{ C,A,E \} }			&&																												\\
																																&&	\mathbb Z_p \ar[ddll]^>>>>>>>{\left(\_ \right)^a,\{ A, E \} }											\\						
																																&&																												\\																																				
 \mathbb Z_p																													&&																												\\
  }
}
\]
whereas the smallest IFO diagram above this is given by replacing the edge labelled $\left[g^{abc}\right] ,\{ A,B,C\} $ with an edge labelled by $\left[g^{abc}\right] ,\{ A,B,C,E\}$. This single change corresponds to the observation that Eve now has a route to calculating Alice, Bob and Carol's shared secret.

By contrast, let us now assume that Eve is in fact aware of Bob's secret key in the $\binom{3}{2}$ version of D-H key exchange.  We modify the diagram of Figure \ref{3-2AE} to take this into account; we replace each ocurrence of $(\ )^b, \{ B \}$ by $(\ ) ^b , \{ B,E \}$ and find the minimal A-E above the result. This gives the diagram of Part {\bf (i)} of Figure \ref{32EknowsB}.  Using the techniques of Section \ref{viewpoints}, we then consider Eve's view of the result, giving part {\bf (ii)} of Figure \ref{32EknowsB}.
\begin{figure}
\caption{When Eve knows Bob's private key}\label{32EknowsB}
\scalebox{0.7}{
\xymatrix{
\mbox{\bf Part (i)}																	&&																																																																																																															&&																																		&&																						&	&	&	&	\mbox{\bf Part (ii)}																&&																																																																																																															&&																																		&&																				\\
																					&&	\mathbb Z_p																																																																																																											&&																																		&&	\mathbb Z_p	\ar[dddd]|{( \ )^c, \{ C \} }	\ar[llll]_{(\ )^a, \{ A \} } 			&	&	&	&																						&&	\mathbb Z_p																																																																																																											&&																																		&&	\mathbb Z_p																 \\
\mathbb Z_p	\ar[dddd]|{(\ )^c,\{ C \} }	\ar[urr]^{(\ )^b, \{ B, E \} }		&&																																																																																																															&&	\mathbb Z_p	\ar[urr]^{( \ )^b, \{ B,E \} } \ar[dddd]|{( \ )^c, \{ C \} } \ar[llll]_<<<<<<<<<<<<<<{(\  )^a, \{ A \} } 	&&																						&	&	&	& \mathbb Z_p								\ar[urr]^{(\ )^b, \{ B,E \} }			&&																																																																																																															&&	\mathbb Z_p	\ar[urr]^{( \ )^b, \{ B,E \} }  																						&&																				\\			
																					&&																																																																																																															&&																																		&&																						&	&	&	& 																						&&																																																																																																															&&																																		&&																					\\ 						 
																					&&																																																																																																															&&																																		&&																						&	&	&	&																						&&																																																																																																															&&																																		&&																						\\ 										 
																					&&	\{ * \}	\ar@[red][uuuu]|!{[uuurr];[uuull]}\hole^{\color{red}[g^{ab}], \{ A , B ,E \} }  \ar@[blue][dll]_{\color{blue}[g^{ca}], \{ C ,A \} }	\ar@[green][rrrr]|!{[rru];[rrd]}\hole ^>>>>>>>>{\color{green}[g^{bc}], \{ B,C,E \} }	\ar[uuurr]|{\color{black}g,\top}	\ar[uuull]|{[g^a],\top }  \ar[drr]|{[g^c],\top } \ar[uuuurrrr]|!{[uuurr];[drr]}\hole _>>>>>>>>>>>>>>>>{[g^b],\top}								&&																																		&&	\mathbb Z_p																		&	&	&	& 																						&&	\{ * \}	\ar@[red][uuuu]|{\color{red}[g^{ab}], \{ A , B, E \} }  	\ar@[green][rrrr]^>>>>>>>>{\color{green}[g^{bc}], \{ B,C,E \} }	\ar[uuurr]|{\color{black}g,\top}	\ar[uuull]|{[g^a],\top }  \ar[drr]|{[g^c],\top } \ar[uuuurrrr] _>>>>>>>>>>>>>>>>{[g^b],\top}																					&&																																		&&	\mathbb Z_p																			\\
\mathbb Z_p																		&&																																																																																																															&&	\mathbb Z_p	\ar[llll]^{( \ )^a, \{ A \} } 	 \ar[urr]	_{( \ )^b, \{ B, E \} }														&&																						&	&	&	& \mathbb Z_p																		&&																																																																																																															&&	\mathbb Z_p	 	 \ar[urr]	_{( \ )^b, \{ B , E \} }																				&&																				\\		
}
}
\end{figure}
It is clear from the diagrams that Eve now has knowledge of the shared secrets of Alice \& Bob, and Bob \& Carol. However, she is not able to discover the shared secret of Alice \& Carol. 

\begin{remark}
Both the above results are course utterly trivial to anyone even slightly familiar with Diffie-Hellman key exchange.  The intention is to demonstrate the reliability of the formalism, before moving on to demonstrate its utility.
\end{remark}

\section{Ambiguity, incompleteness, and algorithmics}\label{incomplete-sect}
In the above diagrammatic manipulations, information about which participant has made a particular announcement is not explicitly included in the A-E diagram for a protocol; this is as described in Section \ref{explain-sect}.  A key point of this formalism is that it nevertheless may be deduced from the context. 

We now move on to situations where we have ambiguous or incomplete information.  
This is not relevant for analysing existing protocols, which are of course carefully designed to avoid ambiguity, and more applicable to real-world situations involving partial information about public and private communications.

\begin{definition}
Let $\mathfrak D$ be a diagram satisfying the IFO condition. 
We say that $\mathfrak D$ is {\bf triangulated} when every non-identity 2-cell is decomposed into composites of identity two-cells, and non-identity two-cells whose source is a path of length two and whose target is a single edge, such as: 
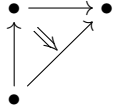
\[ \xymatrix{
\bullet \ar[r]	 \ar@{}[dr]^(.15){}="a"^(.55){}="b" \ar@{=>} "a";"b" & \bullet \\
\bullet \ar[u] \ar[ur] &	\\  }
\]
We say that a {\bf triangulation} of a diagram $\mathfrak D$ is a triangulated diagram $\mathfrak T$ with the same nodes as $\mathfrak D$, that contains $\mathfrak D$ as a sub-diagram. 
\end{definition}
No ambiguity can exist about communication / announcements in a triangulated diagram (beyond the inherent ambiguities given in the original data, such as, `Both $A$ and $B$ know the values $x$ and $y$; one of them subsequently announces the composite $xy$.'). Weaker conditions often suffice to avoid ambiguity. However, for algorithmic purposes the notion of forming triangulations of a given diagram is useful. 

Consider the situation described by the following diagram:
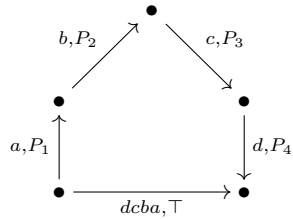
\[ \xymatrix{
												& \bullet  \ar[dr]^{c,P_3}	&					\\
\bullet \ar[ur]^{b,P_2} 								&						& \bullet \ar[d]^{d,P_4} \\
\bullet \ar[u]^{a,P_1}  	\ar[rr]_{dcba,\top}				&						& \bullet 				
}
\]
It is of course inaccurate to declare that, based on the information represented in this diagram, some individual or collection of individuals, in $\bigwedge_{j=1}^4 P_j$ must have publicly announced the result of the composition $dcba$.  A counterexample is given by taking $P_1=\{ V,W\}$, $ P_2=\{ W,X \}$,  $P_3=\{ X,Y \}$, and $P_4=\{ Y,Z \}$,  so $\bigwedge_{j=1}^4 P_j = \bot$.  

Instead, it is clear that when analysing who has shared what information with whom, we require additional edges in that diagram that provide additional {\em epistemic} data but do not add anything to the underlying {\em algebraic} structure.   

Diagrams $\mathfrak D_1$ and $\mathfrak D_2$ below give two possible ways in which the composite $dcba$ came to be public knowledge:
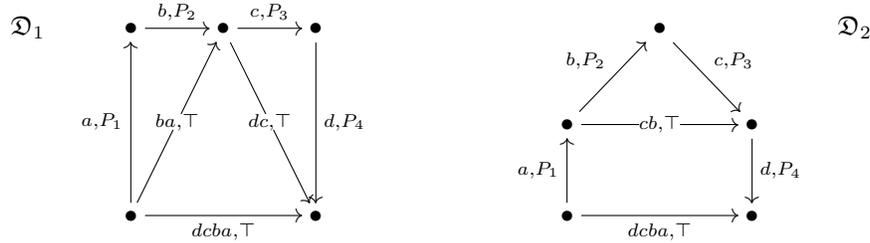
\[ \xymatrix{
\mathfrak D_1		&\bullet \ar[r]^{b,P_2}												& \bullet  \ar[r]^{c,P_3}	\ar[ddr]|{dc,\top}	&\bullet \ar[dd]^{d,P_4} 	&&& 												& \bullet  \ar[dr]^{c,P_3}	&						& \mathfrak D_2 \\
				& 																&									&  						&&&	\bullet \ar[ur]^{b,P_2} 	\ar[rr]|{cb,\top}				&						& \bullet \ar[d]^{d,P_4} 	&				\\
				&\bullet \ar[uu]^{a,P_1}  	\ar[rr]_{dcba,\top}	\ar[uur]|{ba,\top}			&									& \bullet 				 	&&&\bullet \ar[u]^{a,P_1}  	\ar[rr]_{dcba,\top}				&						& \bullet 					&
}
\]
Diagram $\mathfrak D_1$ is triangulated;  we see that $W$ has publicly announced the composite $ba$ and $Z$ has publicly announced $dc$, resulting in any participant being able to compute $dcba$. 

However, $\mathfrak D_2$ is still not triangulated; although we can see that $X$ has publicly announced $cb$ there still remains some ambiguity about how $dcba$ came to be public knowledge. 

To resolve the ambiguity in $\mathcal D_2$, note that it is a sub-diagram of both the following triangulated diagrams: 
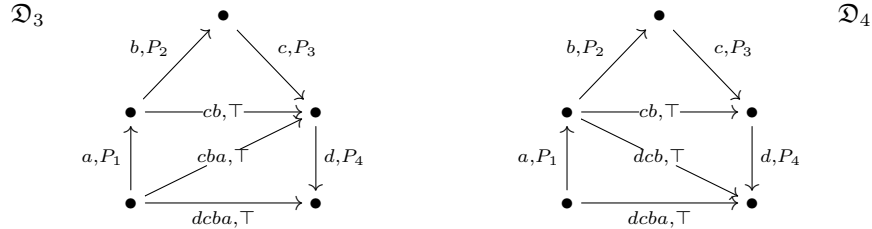
\[ \xymatrix{
\mathfrak D_3	&																				& \bullet  \ar[dr]^{c,P_3}	&								&&& 																		& \bullet  \ar[dr]^{c,P_3}	&						& \mathfrak D_4 \\
				&			\bullet \ar[ur]^{b,P_2} 	\ar[rr]|{cb,\top}								&						& \bullet \ar[d]^{d,P_4}			&&&	\bullet \ar[ur]^{b,P_2} 	\ar[rr]|{cb,\top}	\ar[drr]|{dcb,\top}			&						& \bullet \ar[d]^{d,P_4} 	&				\\
				&		\bullet \ar[u]^{a,P_1}  	\ar[rr]_{dcba,\top} \ar[urr]|{cba,\top}				&						& \bullet 						&&&\bullet \ar[u]^{a,P_1}  	\ar[rr]_{dcba,\top}								&						& \bullet 					&
}
\]
In $\mathfrak D_3$, we see that either $V$ or $W$ has announced $cba$, then either $Y$ or $Z$ has announced $dcba$. Alternatively,    in $\mathfrak D_4$, we see that either $Y$ or $Z$ has announced $dca$ followed by either $U$ or $V$ announcing $dcba$. 

The diagrams $\mathfrak D_1, \mathfrak D_3, \mathfrak D_4$ are of course not the only routes by which $dcba$ may have come to be public knowledge. The two remaining possibilities are left as a straightforward exercise. In general, it is a simple, and easily automated, task to take an A-E diagram and derive the possible ways (if any!) in which communications amongst the participants which may have lead to this situation. 

\begin{remark}
We should be aware that simply drawing such diagrams reflects our own epistemic beliefs; when we tag an edge with the pair $(x,\{U,V\})$ we are making the assumption that, for example, neither $U$ nor $V$ has publicly announced the value $x$. Triangulating a diagram is a method of making deductions about what actions participants may have taken, {\em based on a priori assumptions}.  

For deducing additional information of which we are not aware (e.g. participant $U$ has communicated the value of $x$ to another participant $W$), we must combine the above notion of {\em triangulating diagrams} with the tools derived from considering the poset of diagrams above or below a given diagram.
	\end{remark}

\section{Comparisons \& interactions with other diagrammatic tools}
The intention throughout has been to develop tools that are complementary, rather than competitive, to other graphical or categorical approaches to security.  The objective has been deliberately restricted to the setting where we take the `retrospective' view described in Section \ref{explain-sect}, and use this to reconstruct information flow -- considered generally as both communication, and routes to calculating values.  One perspective is that we are trying to reconstruct, from minimal information, the starting point of a model such as \cite{BGH} that deals with notions of `sites', `channels', and `connections'.

From this viewpoint, it is worthwhile analysing similarities and differences, and potential interactions with other related tools and formalisms.  We consider categorical and graphical settings seperately, although there is of course significant overlap between the two.

The closest approach to this current paper -- from both a categorical and a graphical viewpoint --  is undoubtedly  D. Pavlovic's work `Chasing diagrams in cryptography' \cite{DP}.   The categorical technique of replacing equations by commuting diagrams is both widespread \& powerful, and any category theorist who considers a cryptographic question would naturally start by drawing such diagrams (whether or not they made it into the final work).  

The most substantial difference is that \cite{DP} uses such diagrammatics to reason about (for example) the difficulty of the underlying algebraic problems on which security is based, and (entirely appropriately for this question), the communication between participants and paths they take to solve problems are explicit from the beginning.   It is not hard to imagine some synthesis of his approach and ours, but considering the algebraic problems that must be solved by an attacker would require revisiting the motivation of Footnote 1, that the distinction between a participant and an adversary is sometimes not all that clear.

It is also worth noticing that our notion of `a route to calculating a significant value' is entirely binary, and something that an participant either does, or does not have. In the long term, this needs to become a more fine-grained notion, and the route to calculating a value must be quantified in terms of its cost in terms of time or resources -- something else that is a key concept of \cite{DP}.

A significant precursor to  much modern category theory in security is C. O'Halloran's 1994 DPhil. Thesis, "Category Theory Applied to Information Flow for Computer Security" \cite{COH}.  This brings in yet another strand of category theory into the security world, by finding interpretations of core concepts such as freeness and universal properties. Although not immediately related, it is surely indirectly connected to both this current paper and other category-theoretic works in the field, and is definitely worth revisiting in light of more modern developments in category theory \& security. 

A particular curiosity is that categorical/graphical methods seem more firmly established with regard to quantum-mechanical, rather than classical, protocols.  The starting point for this is undoubtedly the `string diagrams' of Abramsky \& Coecke's `A categorical semantics for quantum protocols' \cite{AC}, along with a great deal of subsequent work.  These `quantum protocols' certainly include  cryptographic protocols\footnote{I would like to thank various members of the Oxford school for the folklore that the `classical communication' in these protocols -- although often implicit -- should properly be thought of as 2-categorical structure. It is pleasing to be able to claim that the same applies to implicit communication in classical protocols!}
    
The use of graphical, rather than specifically categorical, methods in security is much better-established.  It is therefore not possible to give a complete account, but it is still worthwhile to give an overview of particularly popular or similar approaches.  One of the best-established must be the notion of `Attack Trees' \cite{BS} or `Threat Trees' \cite{EA}.  At first sight, these appear completely orthogonal to the approach we take -- they are decidedly goal-oriented, and provide a systematic route achieving this goal by splitting it up into smaller steps, then subdividing these, etc.  This appears to be in stark contrast to our approach where we have not even specified a goal -- rather, we take a retrospective view of `who knows what', and attempt to deduce all possible paths by which this may have occurred.  

 However, taking a more goal-oriented view may become essential as we attempt to scale to larger problems.  In a real-world setting, we are unlikely to be interested in all the details of {\em all possible} scenarios that lead to some A-E diagram; instead we would wish to concentrate on some subset of paths that lead to (for example) a crucial shared secret becoming public knowledge. In the long term, this will require a hierarchical approach, and systematic ways of mapping from {\em models of systems} to {\em models of threats} such as \cite{IP}.

Another notable feature of attack trees is that it is commonly to label non-leaf nodes with logical operators -- conjunction when all subtasks are required in order to achieve a goal, or disjunction when any single one of them will suffice.  This is of course a blunt tool that fails to take into account the cost of achieving a sub-goal, or questions of concurrency \& parallelism.  A more sophisticated approach is given by considering resource-sensitive logics, such as variations on linear logic \cite{HMT}.  More fundamentally, \cite{EJB} describes attack trees themselves as formul\ae\ of  a form of linear logic, and the notion of specialisation as a form of linear implication.

At this stage, we are left in the rather unsatisfactory situation of having a close connection between {\em models}, with no similarly obvious connection between what they model.  The categorical models of quantum protocols given in \cite{AC}are directly based on categorical models of linear logic (precisely, the multiplicative-exponential fragment from Girard's Geometry of Interaction program \cite{GOI1})   found in \cite{SA96}.  This current paper also acknowledges its origins in categorical models of linear logic, although the connection is not as direct.

It is perhaps worth observing that the boolean lattice ordering of participants used throughout this paper is of course a model of a very primitive (boolean) logic, and the crucial partial ordering is the implication of this logic.   The connection is likely to become clearer when a more sophisticated notion of labeling and implication is used.

\section{Future directions}
Although it is visually appealing to be able to draw A-E diagarms for protocols, the intention is also to develop concrete tools.  We have taken the view that they must first be shown to be well-founded, which is why this is a purely theoretical paper. The next question is whether they are both {\em accurate} and {\em useful}.  So far, we have demonstrated that they give the expected (\& indeed, well-established) answer to questions we may pose about communication involving D-H key exchange \& other simple protocols.  

The next step must be to apply the formalism \& tools developed to a wider range of more complex situations, arising from real-world examples, as a step towards validation.  Anything but the simplest cases involve non-trivial algorithmics, so a key part of this will involve automating the types of deductions illustrated in this paper, which is work in progress.  As our tools are designed for deriving implicit knowledge from incomplete information, testing them on real-world examples seems an essential next step. 

\section{Acknowledgements}
I have had the good fortune to encounter several cryptographically-minded category theorists, and category-curious cryptographers. Thanks are due to  Chris Heunen (Edinburgh), Delaram Kahrobaei (York), Dusko Pavlovic (Hawaii), and Noson Yanofsky (New York).  Thanks are also due to Morgan Hines, for finding the regular 3D polyhedron associated with the protocol of Appendix \ref{3dshapes}.

 \bibliographystyle{plain}
\bibliography{crypto_refs}

\appendix

\section{A-E diagrams from non-commutative cryptography}\label{CAKEsection}
Although this paper uses various forms of Diffie-Hellman key exchange as illustrative examples, we emphasise that the techniques are entirely general. To this end, we take a brief diversion, and present A-E diagrams for a family of significantly different cryptographic protocols, from the general field of non-commutative cryptography. 

We do this in order to point out how the techniques for constructing A-E diagrams, their interpretation in terms of information flow, and the correctness criterion, are equally valid in different settings.

A general prescription for public-key protocols in non-commutative cryptography is that of Commuting Action Key Exchange (CAKE), introduced in \cite{CAKE}. Many familiar examples arise from this general prescription. In \cite{CAKE}, the following particular form establishes a shared secret (a member of a given monoid) between the usual two parties, Alice and Bob, as follows:

\begin{definition}[The semigroup CAKE protocol]\label{semiCAKE}
	Given a monoid $M$, {\bf Alice} and {\bf Bob} may come to share a private element $\sigma\in M$ via public communication as follows:
	\begin{enumerate}
		\item Alice and Bob agree on two subsets $A,B\subseteq S$ (their respective {\bf key pools}) that point-wise commute i.e. $ab=ba$ for all $a\in A$ and $b\in B$.
		\item A fixed  {\bf root element} $\gamma\in S$ is agreed upon.
		\item Alice chooses her {\bf private key}, a pair of elements $\alpha_1,\alpha_2\in A$, and publicly broadcasts $\alpha_1\gamma\alpha_2$
		\item Bob chooses his {\bf private key}, a pair of elements $\beta_1,\beta_2\in B$, and publicly broadcasts $\beta_1\gamma\beta_2$.
		\item Alice then computes $\alpha_1 \beta_1 \gamma \beta_2 \alpha_2$ 
		and Bob computes $\beta_1\alpha_1\gamma \alpha_2\beta_2$. By the point-wise commutativity of $A,B\subseteq S$, these are equal, giving Alice and Bob's {\bf shared secret} as $\sigma \ = \ \alpha_1\beta_1\gamma\beta_2\alpha_2 \ = \ \beta_1\alpha_1\gamma \alpha_2\beta_2$.
	\end{enumerate}
	The traditional evesdropper {\bf Eve} is assumed to be party to all communications.
\end{definition}

We treat the monoid $M$ as a category in the usual way, and denote its unique object by $\bullet \in Ob(M)$. The A-E diagram over $M\times 2^{ \{ A,B,E \} }$ for the CAKE family of protocols is given in Figure \ref{lgCAKE}.

\begin{figure}[h]\caption{The Algebraic-Epistemic diagram for semigroup CAKE}\label{lgCAKE}
	{\small 
		\[ 
		\xymatrix{
			&		&  \bullet  \ar[dll]|{\alpha_2, \{ A \} }	 \ar@[red][rrrrr]|{ \sigma , \{ A,B\}  } \ar@{-}[dr]|{\beta_2, \{ B \} }		 						&			&																					& 									&				&  \bullet 													&		&		\\
			\bullet	\ar[ddrr]|{\beta_2,\{ B \} } \ar[rrrrr]|>>>>>>>>>>>>{  P_B , \top  }									&		&																								&	\ar[dr]	&																					&	\bullet	 \ar[urr]|{\alpha_1, \{ A \}}			&				&														&		&		\\
			& 		&																								&			& \bullet \ar[dll]|{ \alpha_2, \{ A \} } \ar@{-}[rr]|{ P_A , \top  } 												&									& \ar[rrr]			& 														&		&  \bullet \ar[uull]|{ \beta_1 , \{ B\} }	\\
			&		& \bullet \ar[rrrrr]|{ \gamma , \top }																			&			&																					&  									&				& \bullet \ar[urr]|{ \alpha_1, \{ A \} }	 \ar[uull]|<<<<<<<{ \beta_1, \{ B\} }			&		&		\\
		}
		\]
	}
\end{figure}

\begin{figure}[h]\caption{Alice and Bob's public CAKE announcements}\label{CAKE_com-fig}
	{\small 
		\[ 
		\xymatrix{
			\bullet	\ar[ddrr]|{\beta_2,\{ B \} } \ar[rrrrr]|>>>>>>>>>>>>{  P_B , \top  }									&		&																								&		&																					&	\bullet	 			&				&														&		&		\\
			& 		&																								&			& \bullet \ar[dll]|{ \alpha_2, \{ A \} } \ar@{-}[rr]|{ P_A , \top  } 												&									& \ar[rrr]			& 														&		&  \bullet 	\\
			&		& \bullet \ar[rrrrr]|{ \gamma , \top }																			&			&																					&  									&				& \bullet \ar[urr]|{ \alpha_1, \{ A \} }	 \ar[uull]|<<<<<<<{ \beta_1, \{ B\} }			&		&		\\
		}
		\]
	}
\end{figure}
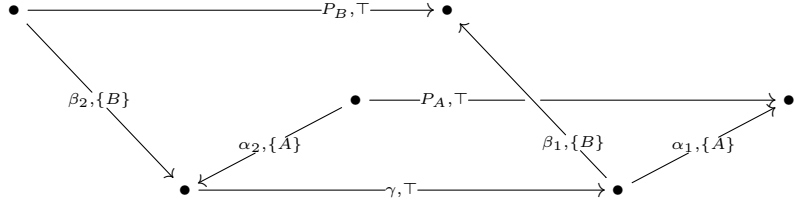

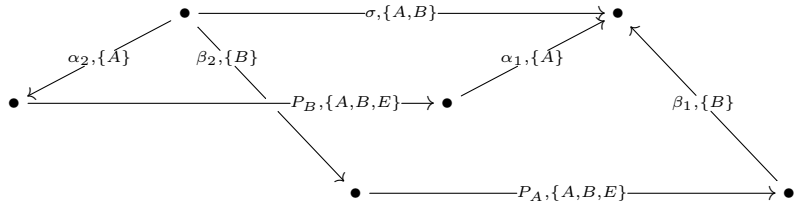
\begin{figure}[h]\caption{Distinct paths to the same result in CAKE}\label{CAKE_path-fig}
{\small 
	\[ \xymatrix{
		&		&  \bullet  \ar[dll]|{\alpha_2, \{ A \} }	 \ar[rrrrr]|{ \sigma , \{ A,B\}  } \ar@{-}[dr]|{\beta_2, \{ B \} } 						&			&																					& 										&				&  \bullet 																				&		&		\\
		\bullet	 \ar[rrrrr]|>>>>>>>>>>>>{  P_B , \{ A,B,E\}  }										&		&																								&	\ar[dr]	&																					&	\bullet	 \ar[urr]|{\alpha_1, \{ A \}}		&				&																					&		&		\\
		& 		&																								&			& \bullet \ar@{->}[rrrrr]|{ P_A , \{ A,B,E\}  } 														&										&		 		& 																					&		&  \bullet \ar[uull]|{ \beta_1 , \{ B\} }	\\
	}
	\]
}
\end{figure}

It is straightforward to verify that the A-E  diagram of Figure \ref{lgCAKE} satisfies the IFO condition.  The non-trivial 2-categorical information (i.e. the two-cells filled in with inequalities) illustrates both public announcements (Figure \ref{CAKE_com-fig}) and distinct routes to calculating the same value (Figure \ref{CAKE_path-fig}).

\begin{remark}[For category-theorists ...] The CAKE family of protocols highlights another, more mysterious, connection between category theory and cryptography. The first algebraic platform proposed, in \cite{SU}, for this family of protocols was Thompson's group $\mathcal F$.  It is by now well-known that Thompson's $\mathcal F$ is precisely the group of {\em canonical associativity isomorphisms}  for a (semi-)monoidal tensor on a monoid, and hence the commuting diagrams representing the algebra of this protocol are canonical diagrams in the sense of MacLane's theory of coherence for associativity \cite{MCL}.  It would be hard to overstate the importance of such diagrams in the foundations of category theory! Even though the protocol of \cite{SU} is well-known to be insecure \cite{FM,RST}, the question remains as to whether this is simply a spectacular coincidence, or an indication of some deeper connection.
\end{remark}

\section{Step-by-step construction of A-E diagrams for $\binom{3}{3}$ DH} \label{stepbystep}
In Section \ref{3-3-DH}, the A-E diagram for the $\binom{3}{3}$ Diffie-Hellman protocol is presented with no indication as to how it was derived. We now give a step-by-step description of the construction of the A-E diagram for the above protocol. 
The algebraic core is the identity $(\_ )^{abc}=(\_ )^{bca}=(\_ )^{cab}$, which we draw as the commuting diagram of Figure \ref{3-3_alg-fig}.
\begin{figure}
	\caption{The algebraic core of $\binom{3}{3}$ D-H key exchange}\label{3-3_alg-fig}
\begin{center}
	\scalebox{0.8}
{
\xymatrix{
						&		&																									& \mathbb Z_p\ar[dd]^{(\_)^c} |!{[dl];[ddr]}\hole		&			\\
\mathbb Z_p\ar[dd]	_{(\_)^b}	&		&  \mathbb Z_p \ar[ur]|<<<<<{(\_ )^b} \ar[drr]|<<<<<{(\_ )^c} \ar[ll]|<<<<<{(\_ )^a}							&												&			\\
						&		&																									&		\mathbb Z_p\ar[dl]|{(\_ )^a}					&\mathbb Z_p 	\ar[dd]^{(\_ )^a}			\\
\mathbb Z_p\ar[rr]|{(\_)^c}	&		& \mathbb Z_p																						&												&			\\
						&		&																									&												&	\mathbb Z_p\ar[ull]	|{(\_ )^b}	\\
  }
}
\end{center}
\end{figure}
The protocol itself relies on these equalities {\em when applied to a specific root element $g\in \mathbb Z$}, so we introduce the singleton object $\{ * \}$ and the element maps $[g],[g^{abc}]: \{ * \} \rightarrow \mathbb Z_p$, giving the commuting diagram of Figure \ref{3-3_DH_root-fig}:
\begin{figure} \caption{The algebra of $\binom{3}{3}$ D-H, applied to a root element} \label{3-3_DH_root-fig}
	\begin{center}
		\scalebox{0.8}{
	$
\xymatrix{
									&&		&																									& \mathbb Z_p\ar[ddddd]^<<<<<<{(\_)^c} |!{[ddl];[dddr]}\hole		&			\\
									&&		&																									& 						 									&			\\
\mathbb Z_p\ar[dddd]_<<<<<<{(\_)^b}	&&		&  \mathbb Z_p \ar[uur]^>>>>>>{(\_ )^b} \ar[drr]^>>>>>>{(\_ )^c} \ar[lll]_>>>>>>{(\_ )^a}					&															&			\\
									&&		&																									&															&\mathbb Z_p 	\ar[dddd]^<<<<<<{(\_ )^a}			\\
									&&		&	\{ * \} \ar[dd]|{\left[g^{abc}\right] } \ar[uu]|{\left[g\right] }												&															&			\\
									&&		&																									&	\mathbb Z_p\ar[dl]|{(\_ )^a}									&								\\						
\mathbb Z_p\ar[rrr]|{(\_)^c}				&&		& \mathbb Z_p																						&															&			\\
									&&		&																									&															&	\mathbb Z_p\ar[ull]	|{(\_ )^b}	\\
  }
$
}
\end{center}
\end{figure}

The elements $g^{a},g^{ab},g^{b},g^{bc},g^{c},g^{ca}$ also play an explicit part in the protocol, so we add in the appropriate arrows from the central point to the outer corners of each vertical rectangle, to give the commuting diagram of Figure \ref{3-3_DH_alg-fig} that describes the interaction of all algebraic entities in $\binom{3}{3}$ Diffie-Hellman key exchange. 
\begin{figure}[h]
	\caption{All the algebraic entities of $\binom{3}{3}$ D-H key exchange}\label{3-3_DH_alg-fig}
	\begin{center}
	\scalebox{0.8}{
		$
\xymatrix{
												&&		&																																																																																											&& \mathbb Z_p\ar[ddddd]^<<<<<<{(\_)^c} |!{[ddll];[dddr]}\hole  |!{[ddddll];[dddr]}\hole		&			\\
												&&		&																																																																																											&& 						 									&			\\
\mathbb Z_p\ar[dddd]_<<<<<<{(\_)^b}	&&		&  \mathbb Z_p \ar[uurr]^>>>>>>{(\_ )^b} \ar[drrr]^>>>>>>{(\_ )^c} \ar[lll]_>>>>>>{(\_ )^a}																																																																	&&															&			\\
												&&		&																																																																																											&&															&\mathbb Z_p 	\ar[dddd]^<<<<<<{(\_ )^a}			\\
												&&		&	\{ * \} \ar[dd]|{\left[g^{abc}\right] } \ar[uu]|{\left[g\right] }\ar[uulll]|{\left[g^{a}\right]} \ar[ddlll]|{\left[g^{ab}\right]}	\ar[rruuuu]|<<<<<<<<<<<<<<<{\left[g^b\right]}|!{[uu];[urrr]}\hole 	\ar[drr]|{\left[ g^{bc}\right]}	\ar[urrr]|<<<<<<{\left[ g^{c}\right]}	\ar[dddrrr]|>>>>>>>>>>>>>>{\left[g^{ca}\right] }	&&															&			\\
												&&		&																																																																																											&&	\mathbb Z_p\ar[dll]^{(\_ )^a}|!{[ull];[ddr]}\hole					&								\\						
\mathbb Z_p\ar[rrr]|{(\_)^c}				&&		& \mathbb Z_p																																																																																							&&															&			\\
												&&		&																																																																																											&&															&	\mathbb Z_p\ar[ulll]	|{(\_ )^b}	\\
  }
$ }
\end{center}
\end{figure}

It finally remains to add in the epistemic data. This is routine, given that 
\begin{enumerate}
\item only Alice (resp. Bob, resp. Carol) can perform $(\ )^a$ (resp. $(\ )^b$, resp. $(\ )^c$).
\item Except for the computations in the bottom triangle,
\begin{itemize} 
\item Alice communicates the results of all her computations to Bob,
\item Bob communicates the results of all his computations to Carol,
\item Carol communicates the results of all her computations to Alice.
\end{itemize}
\item Eve is aware of the results of all communications.
\end{enumerate}
Adding in this information gives the Algebraic-Epistemic diagram of Figure \ref{trinity}; it may be verified that this diagram satisfies the IFO condition, and it is also unambiguous who has communicated what information to whom.
\begin{figure}\caption{Algebraic-Epistemic diagram for $\binom{3}{3}$ Diffie-Hellman}\label{trinity}
\begin{center}
	\scalebox{0.8}
	{
$
\xymatrix{
												&&		&																																																																																																																																					&& \mathbb Z_p\ar[dddddd]^<<<<<<<<<<<{(\_)^c,\{ C \} } |!{[ddll];[dddr]}\hole  |!{[dddddll];[dddr]}\hole			&			\\
												&&		&																																																																																																																																					&& 									 																&			\\
\mathbb Z_p\ar[dddddd]_{(\_)^b,\{ B\} }	&&		&  \mathbb Z_p \ar[uurr]^{(\_ )^b,\{ B\} } \ar[drrr]^>>>>>>{(\_ )^c,\{ C \} } \ar[lll]_{(\_ )^a,\{ A \} }																																																																																																									&&																									&			\\
												&&		&																																																																																																																																					&&																									&\mathbb Z_p 	\ar[dddddd]^{(\_ )^a,\{ A \} }				\\																																				
												&&		&																																																																																																																																					&&																									&		\\
												&&		&	\{ * \} \ar[ddd]|{\left[g^{abc}\right] ,\{ A,B,C\} } \ar[uuu]|>>>>>>>>>{\left[g\right] ,\top}\ar[uuulll]|{\left[g^{a}\right],\{A,B,E\}} \ar[dddlll]|>>>>>>>>>>>>>>{\left[g^{ab}\right],\{ B,C,E \} }	\ar[rruuuuu]|<<<<<<<<<<<<<<<<<{\ \ \ \ \left[g^b\right],\{B,C,E\} }|!{[uuu];[uurrr]}\hole 	\ar[drr]^<<<<<<<{\left[ g^{bc}\right],\{ C,A,E \} }	\ar[uurrr]|<<<<<<<<<<<<{\left[ g^{c}\right], \{ C,A,E \} }	\ar[ddddrrr]|>>>>>>>>>>>>>>{\left[g^{ca}\right], \{ A , B , E \} }	&&																									&			\\
												&&		&																																																																																																																																					&&	\mathbb Z_p \ar[ddll]^>>>>>>>{\left(\_ \right)^a,\{ A \} }|!{[ull];[dddr]}\hole								&								\\						
												&&		&																																																																																																																																					&&																									&			\\																																				
\mathbb Z_p\ar[rrr]_{(\_)^c,\{ C \} }		&&		& \mathbb Z_p																																																																																																																																	&&																									&			\\
												&&		&																																																																																																																																					&&																									&	\mathbb Z_p\ar[ulll]^{(\_ )^b,\{ B \} }	\\
  }
$
}
\end{center}
\end{figure}

\begin{remark}[Generalising to arbitrary  numbers of participants]\label{N-N-DH} It is notable that the diagram of Figure \ref{trinity} consists of three isomorphic diagrams pasted along a common edge, with these three diagrams being related by a cyclic permutation of the symbols $(A,a),(B,b)$ and $(C,c)$.  This clearly generalises to a higher number of participants.
\end{remark}

\section{Diagrammatics for the $\binom{4}{2}$ Diffie Hellman protocol}\label{3dshapes}
As an exercise in diagrammatics, we extend the diagrams of Section \ref{3-3-DH} to the case where there are four participants, and each pair of them establishes a shared secret -- i.e. the $\binom{4}{2}$ D-H key exchange protocol. 
\begin{definition}\label{4-2DH}
	(The {\bf $\binom{4}{2}$ Diffie Hellman protocol})
	We assume participants $\{ {\color{red}Alice}, {\color{green}Bob} ,{\color{blue}Carol},{\color{magenta}Dave}, Eve \}$ where Eve is the evesdropper. Alice, Bob, Carol, and Dave each choose private keys $a,b,c,d\in \mathbb Z_p$, and each pair uses the pair-wise Diffie-Hellman protocol to construct a shared secret.  
\end{definition}

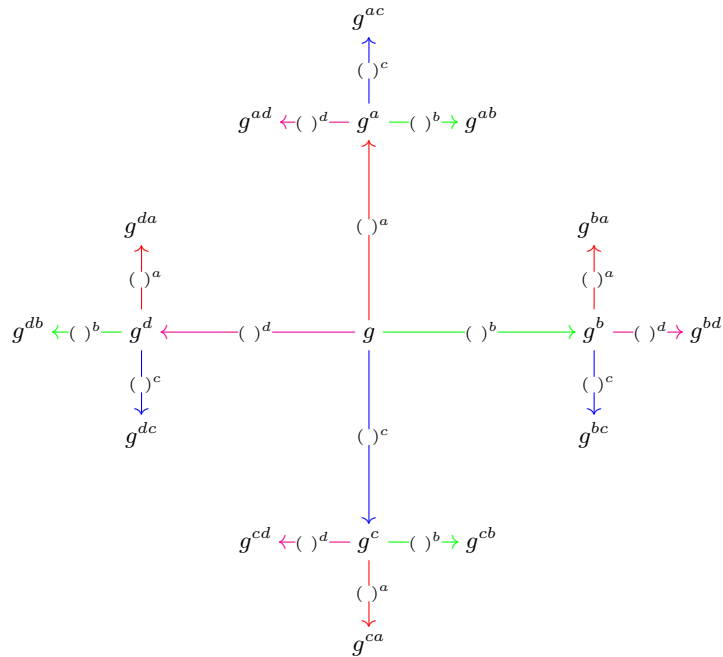
\begin{figure}[h]\caption{Each participant applies their secret operation four times}\label{4-2-graph}
\[ 
\xymatrix{
			&																				&			&	g^{ac}																									&										&																					&			\\	
			&																				&	g^{ad}	&	g^a	\ar@[green][r]|{(\ )^b}	\ar@[blue][u]|{\ (\ )^c} \ar@[magenta][l]|{(\ )^d}							& g^{ab} 								&																					&			\\
			&	g^{da}																		&			&																											&										&	g^{ba}																			&			\\	
	g^{db}	&	g^d	\ar@[red][u]|{\ (\ )^a}\ar@[green][l]|{(\ )^b} \ar@[blue][d]|{\ (\ )^c}	&			&	g \ar@[red][uu]|{\ (\ )^a} \ar@[green][rr]|{(\ )^b} \ar@[blue][dd]|{\ (\ )^c}\ar@[magenta][ll]|{(\ )^d}	&										&	g^b \ar@[red][u]|{\ (\ )^a}	\ar@[blue][d]|{\ (\ )^c} \ar@[magenta][r]|{(\ )^d}	& g^{bd}			\\
			&	g^{dc}																		&			&																											&										&	g^{bc}																			&			\\
			&																				&	g^{cd}	&	g^c	\ar@[red][d]|{\ (\ )^a}	\ar@[green][r]|{(\ )^b} \ar@[magenta][l]|{(\ )^d}							& g^{cb}								&																					&			\\
			&																				&			&	g^{ca}																									&										&																					&			\\		
}
\]
\end{figure}

As a starting point to drawing the A-E diagram for this protocol, let us adopt yet another colour-coding convention, and denote operations applied by each participant by a consistent colour. Each participant then applies the `exponentiation by their private key' on four different occasions, as illustrated in Figure \ref{4-2-graph}.
Recalling commutativity of modular exponentiation operations (i.e. $\left(g^b\right)^c = \left( g^c\right)^b$, \&c.),  we observe that a total of six distinct private keys are calculated, each one in two different ways. 

\begin{figure}
	\caption{Exponentiation by a private key, performed by its owner} \label{4-2-triv}
\[ \xymatrix@C=3em{ 
	&																						&		&\Zp\ar@[magenta][dlll]_{(\ )^d,\{ D \} }	\ar@[green][rrrd]^{(\ )^b,\{ B \} }\ar@[blue][dddl]_{(\ )^c, \{ C \} }													&																																		&		&		&		\\	
	\Zp		&																						&		&																																		&																																		&		&\Zp	&		\\	
	&																						&		&																																		&																																		&		&		&		\\	
	&																						&\Zp 	&																																		& \Zp \ar@[red][uuul]_(0.25){( \ )^a, \{ A \} }\ar@[green][drrr]^(0.25){(\ )^b, \{ B \} } \ar@[magenta][dlll]_(0.25){(\ )^d, \{ D \} } \ar@[blue][dddl]^(0.25){(\ )^c,\{ C\} }|!{[dlll];[dd]}\hole	&		&		&		\\	
	& \Zp\ar@[red][luuu]^{( \ )^a,\{ A \} } \ar@[blue][lddd]_{( \ )^c,\{ C \} } \ar@[green][rrrd]_(0.25){(\ )^b,\{ B \} } 	&		&																																		&																																		&		&		&\Zp\ar@[red][luuu]_{( \ )^a, \{ A \} }	\ar@[blue][dddl]^{(\ )^c, \{ C \} } \ar@[magenta][dlll]^{(\ )^d, \{ D \} } \\	
	&																						&		&																																		&	\Zp																																	&		&		&		\\	
	&																						&		& \Zp\ar@[magenta][dlll]^{(\ )^d,\{ D \} } \ar@[green][rrrd]_{(\ )^b,\{ B\} } \ar@[red][luuu]^(0.25){( \ )^a, \{ A \} } |!{[lluu];[ur]}\hole |!{[lluu];[uuur]}\hole		&																																		&		&		&		\\	
	\Zp		&																						&		&		&		&		&\Zp	&		\\	
}
\]
\end{figure}
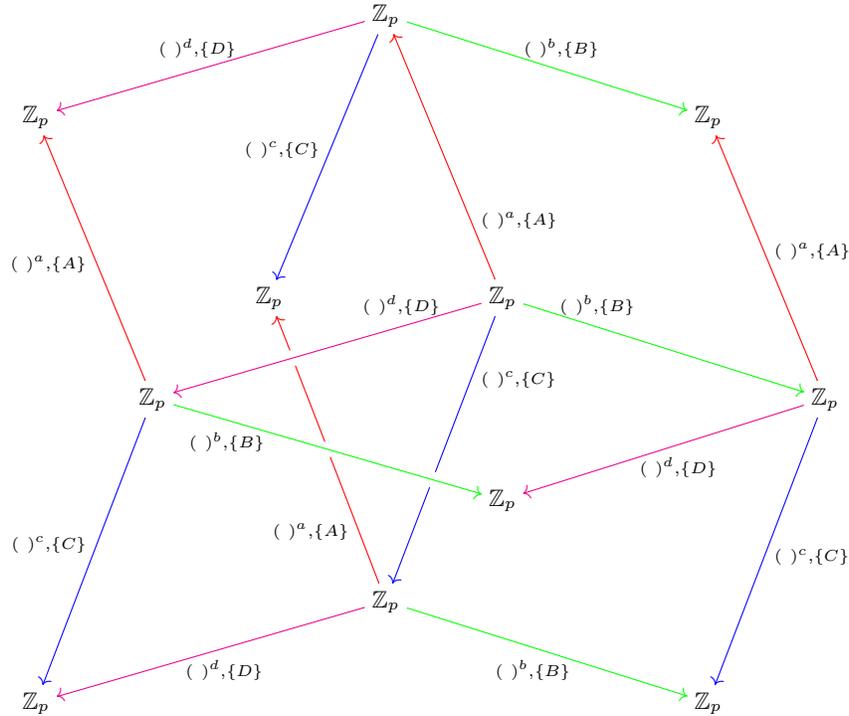

Let us now take the graph of Figure \ref{4-2-graph} and identify nodes that have the same value (e.g. the top $g^{ca}$ and the bottom $g^{ac}$). We may draw the resulting figure as a regular three-dimensional figure where all lines have equal length, and all edges with the same colour are parallel. As these identifications are valid regardless of the value of the root $g\in \Zp$, we have replaced the individual values by the object $\Zp$. We have also added in the epistemic operation -- this step is trivial, since as each operation shown(i.e. exponentiation by a private key) this may only be performed by the respective owner. We derive the (commuting) diagram of Figure \ref{4-2-triv}. 

Note that this diagram does indeed commute -- there are no non-trivial 2-cells. The composite along any two paths with the same source and target is of the form $(\ )^s,\{ \}$, for some shared secret $s\in \mathbb Z_p$.  This emphasises that no single individual is able to calculate a shared secret without a public announcement from another participant! Adding in the key singleton object, and the respective `select an element' arrows will provide for non-trivial 2-cells, and hence models of knowledge \&  information flow. 

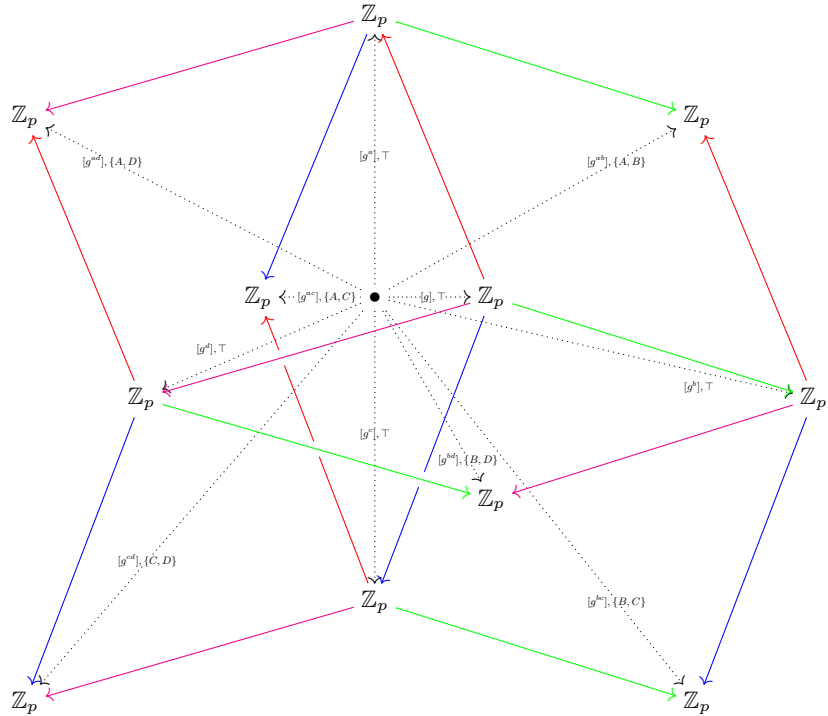
\begin{figure}
	\caption{The A-E diagram for $\binom{4}{2}$ Diffie-Hellman}\label{4-2-AE}
\[ \xymatrix@C=3em{ 			
		&																														&		&\Zp\ar@[magenta][dlll]	\ar@[green][rrrd]\ar@[blue][dddl]																												&																								&		&		&		\\	
		\Zp		&																														&		&																																										&																								&		&\Zp	&		\\	
		&																														&		&																																										&																								&		&		&		\\	
		&																														&\Zp 	&	\bullet		 \ar@{.>}[l]|(0.41){\scalebox{0.45}{ $[g^{ac}],\{ A ,C \}$ }} \ar@{.>}[llluu]|(0.75){\scalebox{0.45}{ $[g^{ad}],\{ A ,D \}$ }} \ar@{.>}[uuu]|{\scalebox{0.45}{$[g^a],\top$}} \ar@{.>}[rrruu]|(0.75){\scalebox{0.45}{$[g^{ab}],\{A,B\}$} } \ar@{.>}[r]|{\scalebox{0.45}{$[g],\top$}} \ar@{.>}[rrrrd]_(0.75){\scalebox{0.45}{$ [g^b],\top $}} \ar@{.>}[rrrdddd]|(0.75){\scalebox{0.45}{ $[g^{bc}],\{ B ,C \}$ }} \ar@{.>}[rdd]|(0.8){\scalebox{0.45}{ $[g^{bd}],\{ B ,D \}$ }} 	 \ar@{.>}[ddd]|(0.45){\scalebox{0.45}{ $[g^{c}],\top$ }} \ar@{.>}[ddddlll]|(0.65){\scalebox{0.45}{ $[g^{cd}],\{ C,D \}$ }}	
		\ar@{.>}[dll]_(0.65){\scalebox{0.45}{ $[g^{d}],\top$ }}								& \Zp \ar@[red][uuul]\ar@[green][drrr] \ar@[magenta][dlll] \ar@[blue][dddl]|!{[dlll];[dd]}\hole	&		&		&		\\	
		& \Zp\ar@[red][luuu] \ar@[blue][lddd] \ar@[green][rrrd]																	&		&																																										&																								&		&		&\Zp\ar@[red][luuu]	\ar@[blue][dddl] \ar@[magenta][dlll] \\	
		&																														&		&																																										&	\Zp																							&		&		&		\\	
		&																														&		& \Zp\ar@[magenta][dlll] \ar@[green][rrrd] \ar@[red][luuu] |!{[lluu];[ur]}\hole |!{[lluu];[uuur]}\hole																	&																								&		&		&		\\	
		\Zp		&																														&		&																																										&																								&		&\Zp	&		\\	
}
\]
\end{figure}

Unfortunately, three spacial dimensions do not suffice for drawing this as a {\em regular} shape;  we therefore adopt the following simplifications in order to make our diagram manageable:
\begin{itemize}
	\item As both the algebraic and epistemic labels on the coloured edges are uniquely determined by the colour, we omit these labels in favour of the colour-coding only. 
	\item The `select an element' arrows are drawn as dotted lines, with no attempt made to distinguish over- and under- crossings.
	\item The unique singleton element is represented by a bullet $\bullet=\{ * \}$.
\end{itemize}
This then leads to the diagram shown in Figure \ref{4-2-AE}, which may be verified to satisfy the IFO condition.

\begin{remark}[Combinatorics, diagrams, and polyhedra for $\binom{n}{k}$ Diffie-Hellman] \\
Combinatorially, it is relatively easy to write down an abstract characterisation -- in terms of nodes, edges, and labels -- of the A-E diagram for the $\binom{n}{k}$ Diffie-Hellman protocol, where there are $n$ participants, and every subgroup of $k$ individuals comes to share a secret. What is less intuitive is the description of these diagrams in terms of regular figures in $d$-dimensional space.  A suitable geometric characterisation surely exists, although it is far from the stated aims of this paper! It is therefore left as a potentially non-trivial exercise.
\end{remark}

\end{document}